# Bayesian data assimilation for estimating epidemic evolution: a COVID-19 study

Xian Yang[1,2,¶], Shuo Wang[2,3,4,¶], Yuting Xing[2,¶], Ling Li[5], Richard Yi Da Xu[6], Karl J. Friston[7] and Yike Guo[1,2,*]

[1]Department of Computer Science, Hong Kong Baptist University, Hong Kong Special Administrative Region, China

[2]Data Science Institute, Imperial College London, the United Kingdom

[3]Digital Medical Research Center, School of Basic Medical Sciences, Fudan University, China

[4]Shanghai Key Laboratory of Medical Image Computing and Computer Assisted Intervention, China

[5]School of Computing, University of Kent, the United Kingdom

[6]Faculty of Engineering and Information Technology, University of Technology Sydney, Australia

[7]Institute of Neurology, University College London, the United Kingdom

[¶]These authors contributed equally to this work.

[*] Corresponding author:

E-mail: yikeguo@hkbu.edu.hk

## Abstract

The evolution of epidemiological parameters, such as instantaneous reproduction number $R_t$, is important for understanding the transmission dynamics of infectious diseases. Current estimates of time-varying epidemiological parameters often face problems such as lagging observations, averaging inference, and improper quantification of uncertainties. To address these problems, we propose a Bayesian data assimilation framework for time-varying parameter estimation. Specifically, this framework is applied to $R_t$ estimation, resulting in the state-of-the-art 'DARt' system. With DARt, time misalignment caused by lagging observations is tackled by incorporating observation delays into the joint inference of infections and $R_t$; the drawback of averaging is overcome by instantaneously updating upon new observations and developing a model selection mechanism that captures abrupt changes; the uncertainty is quantified and reduced by employing Bayesian smoothing. We validate the performance of DARt and demonstrate its power in revealing the transmission dynamics of COVID-19. The proposed approach provides a promising solution for accurate and timely estimating transmission dynamics from reported data.

## Author Summary

Monitoring the evolution of transmission dynamics is of great importance in response to the COVID-19 pandemic. The transmission dynamics of infectious disease is described by epidemiological models, but the model parameters may vary substantially due to differences in government intervention policies. Existing methods on estimating time-varying epidemiological parameters face problems such as lagging observation, averaging inference, and unreliable uncertainty. To address these issues, we have proposed the Bayesian data framework to provide a timely estimate with credibility interval. We have developed the 'DARt' system to monitor the instantaneous reproduction number $R_t$ from daily COVID-19 reports. The accuracy and robustness of our system are validated in numerical simulations and in retrospective analysis of real-world scenarios. Our system provides the insights of impacts of different intervention polices and highlights the effectiveness of undergoing mass vaccination.

## Introduction

Epidemic modelling is important for understanding the transmission dynamics and responding to the emerging COVID-19 pandemic[1]–[8]. Since the pilot work by Kermack and McKendrick[9], various epidemic models with different governing equations have been developed to describe the transmission dynamics of infectious diseases[10]. For common infection diseases such as influenza, the epidemiological parameters are related to the nature of the virus and treated as constants during the epidemic outbreak. These models are not applicable to the emerging COVID-19 pandemic where extensive government control measures have been implemented and revised. Due to the impacts of control measures, the epidemiological parameters (e.g., infection rates) linked to human behaviours could change substantially. In particular, the instantaneous reproduction number $R_t$ has drawn extensive attention, defined as the expected number of secondary cases occurring at time $t$, divided by the number of infected cases scaled by their relative infectiousness[11]. Estimating such time-varying parameters from epidemiological observations (e.g., daily report of confirmed cases) is useful for nowcasting transmission[12], for retrospectively assessing intervention impacts and for developing vaccine strategies[6], [13], [14]. All the applications depend on a reliable system estimating the time-varying parameters with accuracy and timeliness. Imprecise estimation or inappropriate interpretation could feed misinformation. Several systems [10], [15]–[18] have been proposed to estimate the time-varying epidemiological parameters in practice; however, this remains a challenging task due to the following issues[19]:

a. **Lagging observations.** Given a mathematic model of transmission dynamics, to infer the time-varying parameters, the number of infections should be the ideal observable data.

However, the actual infection number is unknown and can only be inferred from other epidemiological observations (e.g., the daily confirmed cases). Such observations are lagging behind the infection events due to inevitable time delays between the infection of individual patients and the detection of the cases (e.g., days for symptom onset[20]). Direct parameter estimation from lagging observations without adjusting for the time delay results in the temporal inaccuracy of estimates[19], [21]. To address this problem, a two-step strategy, first estimating infections from epidemiological observations with a temporal transformation followed by parameter estimation, has been commonly used in practice[21]. The simple temporal shift of observations by the mean observation delay turns out not sufficient for the relatively long observation delay or the rapidly changing transmission dynamics, which are seen in the COVID-19 pandemic[21]. Backward convolution method (i.e., subtracting time delay, with a given distribution, from each observation time) leads to an over-smooth reconstruction of the infection number and bias for parameter estimation[10]. Deconvolution methods[22] through inversing the observation process are mathematically more accurate but sensitive to the optimisation procedures (e.g., stopping criterion) of the ill-posed inverse problem. In addition, the estimated result of infection number is often calculated as a point estimate so that the uncertainty from the observation process is neglected[23]. Taking an alternative method to the two-step strategy, we are investigating a new Bayesian approach that could jointly estimate both infection number and epidemiological parameters with uncertainty by explicitly parameterising the observation delay.

b. **Averaging inference.** There are two general paradigms to deal with the challenge of estimating time-varying parameters: 1) reformulating the problem into an inference of static or quasi-static parameters, so that various methods for static parameter estimation can be used; 2) developing inference methods for explicit time-varying parameter estimation. For the first approach, the time-varying parameter is usually parameterised with several static parameters (e.g., the initial value and the exponential decay rate[17]). The quasi-static method is to assume such a slow evolution of the parameter that could be treated as static within a short period. For example, Cori et al.[16] proposed a sliding-window method 'EpiEstim' using a segment of observations for the averaging inference of $R_t$, assuming $R_t$ remains the same within the sliding window. But this assumption does not apply to the rapidly changing transmission dynamics, and the window size affects the accuracy. Best practices of selecting the sliding window are still under investigation[21]. Instead of adopting a local sliding window, Flaxman et al.[13] defined several periods according to the dates of intervention measures, assuming a constant $R_t$ within each period. This approach requires additional information about the intervention timeline, which could be inaccurate, and does not capture the abrupt change of $R_t$. In contrast to these window-based methods, data assimilation[24] is a window-free alternative approach that has been less explored for parameter estimation in computational epidemiology. Applying sequential Bayesian inference[25], [26], data assimilation supports instantaneous updating of model states upon the arrival of new observations. The Bayesian model selection mechanism[27] can also be used for modelling the switching transmission dynamics under interventions, avoiding the drawback of averaging inference. Different from the common compartment models[28]–[30] used in concurrent data assimilation studies of COVID-19 modelling, we use the renewal process taking account for the changing infectiousness of the virus during the infection period. Moreover, we propose the Bayesian smoothing scheme that allows the correction of historical estimate based on subsequent observations.

c. **Quantification of uncertainty.** The credibility of parameter estimation is of equal importance compared to the estimate itself, especially for policymaking. The uncertainty come from different sources, including the intrinsic uncertainties of epidemic modelling, data observation and inference processes. Firstly, the uncertainty of epidemiological models affects the final estimates. For example, $R_t$ estimation is found to be sensitive to the assumed distribution of generation time intervals[21]. Secondly, the uncertainty, resulting from systematic errors (e.g., weekend misreporting) and random errors (e.g., spike noise) in the observation processes should be properly quantified. During the COVID-19 pandemic, for example, we have seen different reporting standards and time delay across countries and regions, with different levels of uncertainty. Thirdly, the uncertainty could be enlarged or smoothed in the inference processes. For example, the use of a sliding window could smooth the parameter estimation but may simultaneously miscalculate the uncertainty, due to the overfitting within the sliding-window. To provide reliable credible intervals (CrI) of parameter estimates, the three aforementioned types of uncertainty should all be considered and reported as part of the final estimates. As a state-of-the-art package for $R_t$ estimation, EpiEstim (Version 2)[31] allows users to account for the uncertainty from epidemiological parameters by resampling over a range of plausible values. However, the uncertainty from imperfect observations and the side effects associated with the sliding window cannot be processed by this tool. Recently, 'EpiNow'[18] was proposed to integrate the uncertainty of observation process, but the inference is still based on the sliding window. In this work, we deal with model and data uncertainty in the data assimilation framework[24] with a Bayesian smoothing mechanism to enable both the latest and historical observations to be continuously integrated into inference flow, thereby alleviating spurious variability of estimations.

In order to tackle these practical issues, we propose a comprehensive Bayesian data assimilation system, for estimating time-varying epidemiological parameters together with their uncertainty. In particular, we focus on the joint estimation of infection numbers and $R_t$ as a real-world application. Compared to the Bayesian approach for estimating the basic reproduction number $R_0$ at the beginning stage of an epidemic break[32], the sequential updating scheme is developed in our system. The evolution of the transmission dynamics is described by a hierarchical transition process, which is informed by newly data formulated with explicit observation delay. A model selection mechanism is built in the transition process to detect abrupt changes under interventions.

# Results

## 1. Bayesian Data Assimilation for Epidemiological Parameter Estimation

We propose a Bayesian data assimilation approach, as illustrated in Figure 1, to estimate the time-varying parameters based on epidemiological observations. This framework is applicable to various epidemic models when the governing equations and observation functions are available. Given an epidemic model (e.g., renewal process), we can construct a latent state $\boldsymbol{X}_t$ at time $t$ which consists of the time-varying variables and parameters of the governing equations. The epidemiological observations $C_{1:T}$ up to the latest observation time $T$ are made during the observation process of the latent state $\boldsymbol{X}_t$. The problem of estimating the time-varying parameters can be formulated as a Bayesian inference problem of $p(\boldsymbol{X}_t|C_{1:T})$ for each time step $t$. In contrast to inferring the 'pseudo' dynamics (i.e., reformulating into a static/quasi-static problem), our method directly estimates the 'real' dynamics by assimilating information from the observations for the epidemic model forecast. Please refer to the Method section for more detailed explanations.

## (A) Forward filtering at each time step

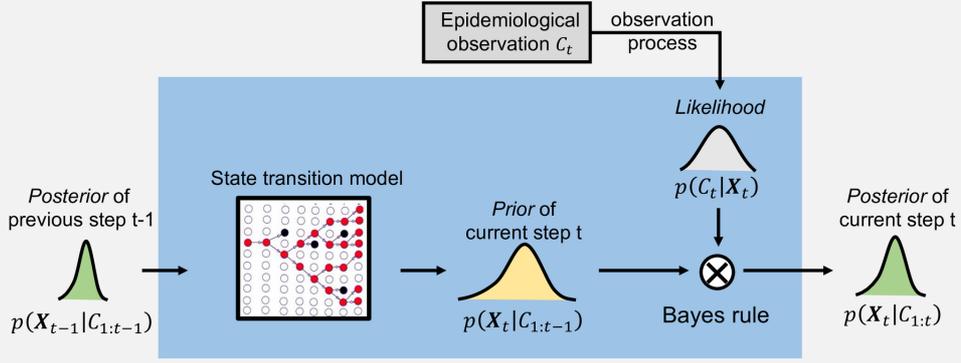

## (B) Forward filtering and backward smoothing

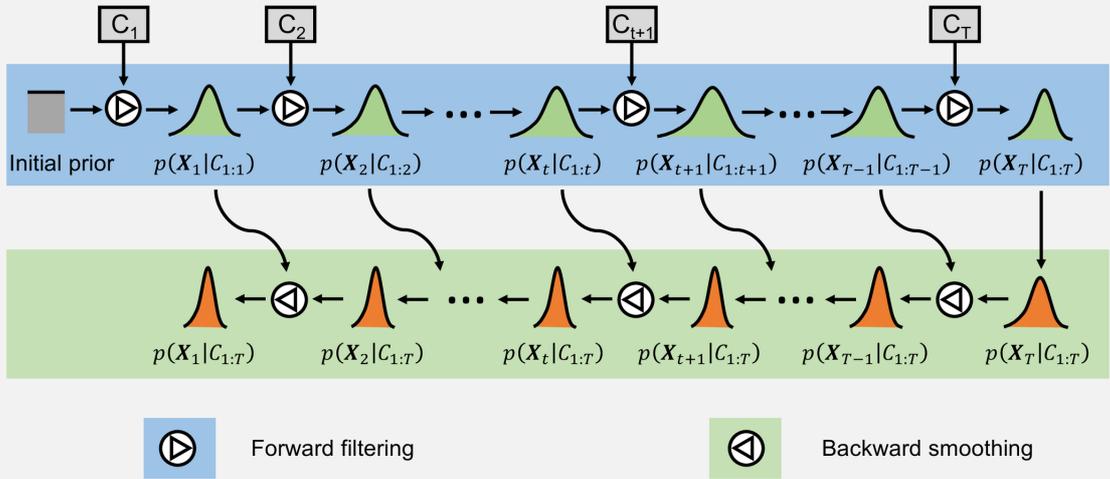

**Figure 1.** Illustration of the inference of Bayesian data assimilation system for time-varying parameter estimation. The latent state $X_t$ includes the variables and parameters of an epidemic model to be estimated. The epidemiological observation is denoted as $C_t$, and is linked to the latent state via the observation function. For each time step, the estimation of the latent state $p(X_t|C_{1:t})$ is constantly updated according to ongoing reported observations using sequential Bayesian updating with forward filtering and backward smoothing. **(A) Forward filtering at each time step**. The posterior state estimation $p(X_{t-1}|C_{1:t-1})$ estimated from previous step $t-1$ is transformed as the *prior* $p(X_t|C_{1:t-1})$ for the current step $t$, calculated from the state transition model as detailed in the Method section. Together with the *likelihood* $p(C_t|X_t)$ obtained from epidemiological observation at the current step, the *posterior* of the current step $p(X_t|C_{1:t})$ is estimated. At the same time, as shown in **(B), backward smoothing** is used to compute $\{p(X_t|C_{1:T})\}_{t=1}^{T}$, taking account of all the observations $C_{1:T}$ up to the time $T$

by applying a Bayesian smoothing method (see the Methods section for more informaiton).

## 2. DARt: A Data Assimilation System for $R_t$ Estimation

To apply the proposed Bayesian data assimilation approach in a real-world problem, we developed the 'DARt' (Data Assimilation for $R_t$ estimation) system for the $R_t$ estimation. The transmission dynamics is described by the governing equations of the renewal process, where $R_t$ is the key epidemiological parameter driving the number of incident infections $j_t$. We construct the latent state $\boldsymbol{X}_t$ including the variable $j_t$, the time-varying parameter $R_t$ and the auxiliary variable $M_t$ for switching dynamics. The evolution of the latent state $\boldsymbol{X}_t$ can be described using a hierarchical transition model, as detailed in the Methods section. Under the modelling of convolutional observation process, we test the capacity of DARt with different observation inputs and kernels.

The performance of the DARt system is validated and compared to that of the state-of-the-art EpiEstim and EpiNow2 systems through simulations and real-world applications. The results confirm its power of estimation and adequacy for practical use. We have made the system available online for broad use in $R_t$ estimation for both research and policy assessment.

- **Validation through simulation**

Due to the lack of ground-truth $R_t$ in real-world epidemics, we conduct a set of simulation experiments by using synthetic data for validation. Figure 2 illustrates the design of simulation experiments where a synthesised $R_t$, is adopted as the ground truth to validate its estimated $\hat{R}_t$. We also estimated $R_t$ using the state-of-the-art $R_t$ estimation package EpiEstim[31] and EpiNow2[18] to compare the effectiveness in overcoming the three aforementioned issues (i.e., lagging, averaging and uncertainty).

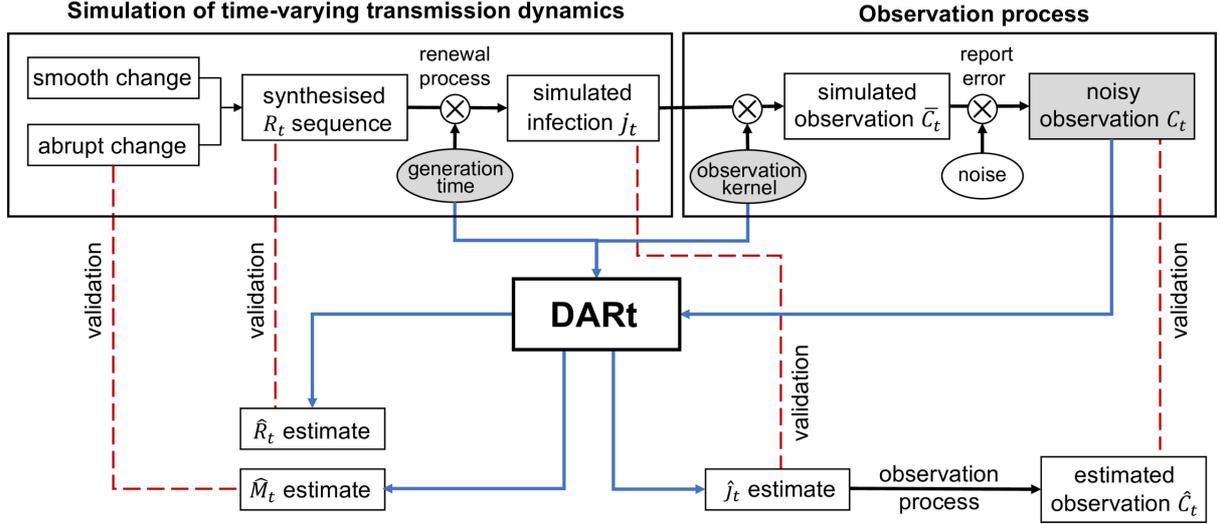

**Figure 2.** Validation experiment of the DARt system on simulated data. First, the ground-truth $R_t$ sequence is synthesised using piecewise Gaussian random walk split by several abrupt change points. The sequence of incident infection $j_t$ is simulated based on a renewal process parameterised by the synthesised $R_t$. The observation process includes applying a convolution kernel that represents the probabilistic observation delay to obtain the expected observation $\bar{C}_t$ and adding Gaussian noise that represents the reporting error to obtain the noisy 'real' observation $C_t$. The inputs (in grey) to the DARt system are the distributions of generation time, observation kernel and simulated noisy observation $C_t$. The system outputs are the estimated $\hat{R}_t$, estimated $\hat{j}_t$ and change indicator $\hat{M}_t$. These outputs are compared with the synthesised $R_t$, $j_t$ and the time of abrupt changes. Also, the observation function is applied to the estimated $\hat{j}_t$ to compute the estimated observation $\hat{C}_t$ with uncertainty, which is compared to the 'real' observation.

**Experimental settings.** In the simulation experiments, we compare the performance of DARt with that of two comparative methods: EpiEstim and EpiNow2. These two models are applied under their default settings with a 7-day smoothing window. EpiEstim takes observations $C_t$ as the infection without adjusting for the observation delay, which is a 'plug-and-play' use of EpiEstim as reported in [1]. As the current implementation of EpiNow2[1] only supports Gamma and Lognormal distribution as time delays, we set the generation time distribution to be a Gamma distribution with shape and scale equal to 4.44 and 1.89 (obtained by fitting the Weibull distribution reported by Ferretti et al.[3] using the Gamma distribution). With the simulated $R_t$ curve and the generation time distribution, we follow the renewal process to simulate the infected curve $j_t$ (initialized to be 1). Then, the observation curve of onset cases $\bar{C}_t$ is generated using the incubation time distribution[3] (i.e., the lognormal distribution with log mean and standard deviation of 1.644 and 0.363 days respectively) as the observation time delay. Similar to the experiments in other related work [12], [13], all comparative models start estimation when the daily observation exceeds a threshold number, which is set to be 10 in our experiments.

---
[1] https://github.com/epiforecasts/EpiNow2

Figure 3 (A) shows the synthesised $R_t$ curve following a piece-wise Gaussian random walk that mimics the scenario of two successive interventions and one resurgence. To approximate the early stage of exponential growth, the simulation starts with $R_0 = 3.2$ (i.e. the basic reproduction number) and follows a Gaussian random walk $R_{t+1} \sim Gaussian(R_t, (0.05)^2)$. At $t = 23$, we set $R_{23}=1.6$ indicating the mitigation outcome of soft interventions. After soft interventions, the epidemic is still being uncontrolled with the evolution of $R_t$ resuming to the Gaussian random walk as above. At $t = 33$, $R_t$ experiences another abrupt decrease to a value under 1, where we set $R_{33}=0.5$ to indicate the suppression effects of intensive interventions (e.g., lockdown). After the epidemic is controlled for a while, one outbreak happens at $t = 83$ with $R_{83}=3$. The evolution of $R_t$ after this resurgence follows the random walk for a few days.

To simulate the real-world noisy observations, we added Gaussian noise with zero mean and standard deviation equal to $N$ times of $\bar{C}_t$. The results presented in the following parts of the main manuscript are obtained with $N = 1$. To further investigate the performance of all comparable models under different levels of noise, we also show the results when $N$ is chosen from {0,1,2,3} in Supplementary Figure 6. Notably, in the rest of this main manuscript both the generation and incubation time distributions are truncated and normalised, where values smaller than 0.1 are discarded. Sensitivity analysis has been done in Supplementary Figure 7, showing impacts of different choice of threshold. Supplementary Figure 8 further investigates the uncertainties resulted from different settings of time distributions.

**Simulation results**

All simulation results are shown in Figure 3 and our observations are summarized as follows.

- **Correctness of $R_t$ estimation**: Figure 3 (B) compares the synthesised $R_t$ with the estimated $R_t$ from DARt, EpiEstim and Epinow2. We can see that $R_t$ from DARt matches the synthesised $R_t$ better than that from the other comparative methods with relatively less degree of fluctuations and faster response to abrupt changes. The results demonstrate that the proposed model can mitigate the influence of noisy observations and overcome the weakness of averaging. The probabilities of having abrupt changes are captured by $M_t$ as shown in Figure 3 (C). Even under observation noise, DARt can still detect abrupt changes.

- **Correctness of $j_t$ estimation**: Figure 3 (D) shows the simulated $j_t$, DARt estimated $j_t$, and EpiNow2 estimated $j_t$. We can find that the DARt estimated $j_t$ with 95% CrI match well the simulated $j_t$. In contrast, the $j_t$ curve from Epinow2 does not align with the simulated $j_t$ very well. In particular, the peak value of $j_t$ from EpiNow2 deviates greatly from the simulated value.

- **Accuracy in recovering observations $C_t$**: Figure 3 (E) compares the distributions of reconstructed $C_t$ from DARt and that of EpiNow2. We can find that compared with $C_t$ from EpiNow2, $C_t$ from DARt with 95% CrI can generally match well with the simulated $C_t$.

- **Effectiveness of DARt smoothing**: Figure 3 (F) illustrates the effectiveness of backward smoothing by comparing the DARt estimated $R_t$ results with and without smoothing, showing the expected smoothing effect of estimated $R_t$ with reduced CrI. It is clear that the results from DARt without smoothing are affected by local fluctuations, which are probably due to observation noises. With the introduction of smoothing, both the uncertainties and local fluctuations are reduced.

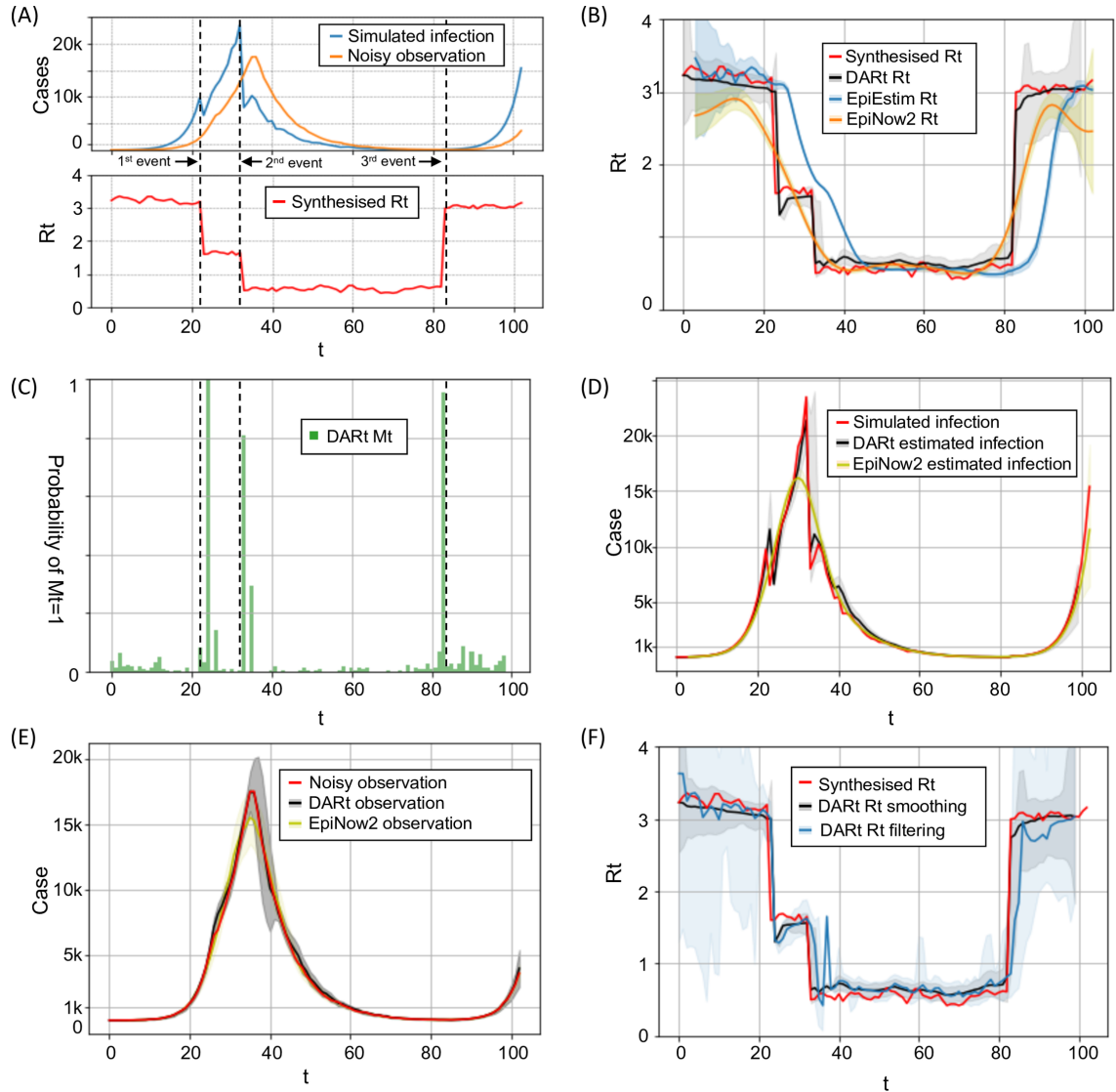

**Figure 3. Simulation results.** **(A)** Synthesised $R_t$, simulated $j_t$ and $C_t$ curves. **(B)** shows the comparison of the synthesised $R_t$ (in red) with estimated $R_t$ curves from DARt, EpiEstim and EpiNow2. **(C)** shows the estimated $M_t$ from DARt to indicate sharp changes of $R_t$. **(D)** shows the simulated $j_t$, $j_t$ from DART, and $j_t$ from EpiNow2. **(E)** compares the distributions of estimated $C_t$ from DARt and EpiNow2 with the simulated $C_t$ curve with 95% CrI. **(F)** compares the DARt estimated $R_t$ results with and without smoothing.

- **Applicability to real-world data**

We applied DARt to estimate $R_t$ in four different regions during the emerging pandemic. Each region represents a distinct epidemic dynamic, allowing us to test the effectiveness and robustness of DARt in each scenario. **1) Wuhan**: When the outbreak of COVID-19 happened in Wuhan, the government responded with very stringent interventions such as a total lockdown. By studying its $R_t$ evolution, we can check the capability of DARt in detecting the abrupt changes of $R_t$. **2) Hong Kong**: The daily increase of reported cases in Hong Kong has been remained at a low level for most of the time with the maximum daily value under 200. As no stringent interventions have been introduced in such a city with high-density population, Hong Kong offers an ideal scenario for studying the change of $R_t$ under mild interventions. **3) United Kingdom:** The daily infection number in the UK changed significantly this year, varying from 2,000 to 6,000. UK is one of the first countries initiating mass immunisation campaign; therefore, its instantaneous $R_t$ is a useful metric for checking the utility of vaccine in real world on the way towards 'herd immunity'. **4) Sweden:** Sweden is a representative of countries that have less stringent intervention policies; it has a clear miss-reporting pattern repeated periodically. This makes Sweden an ideal case to examine the robustness of DARt with considerable observation noises.

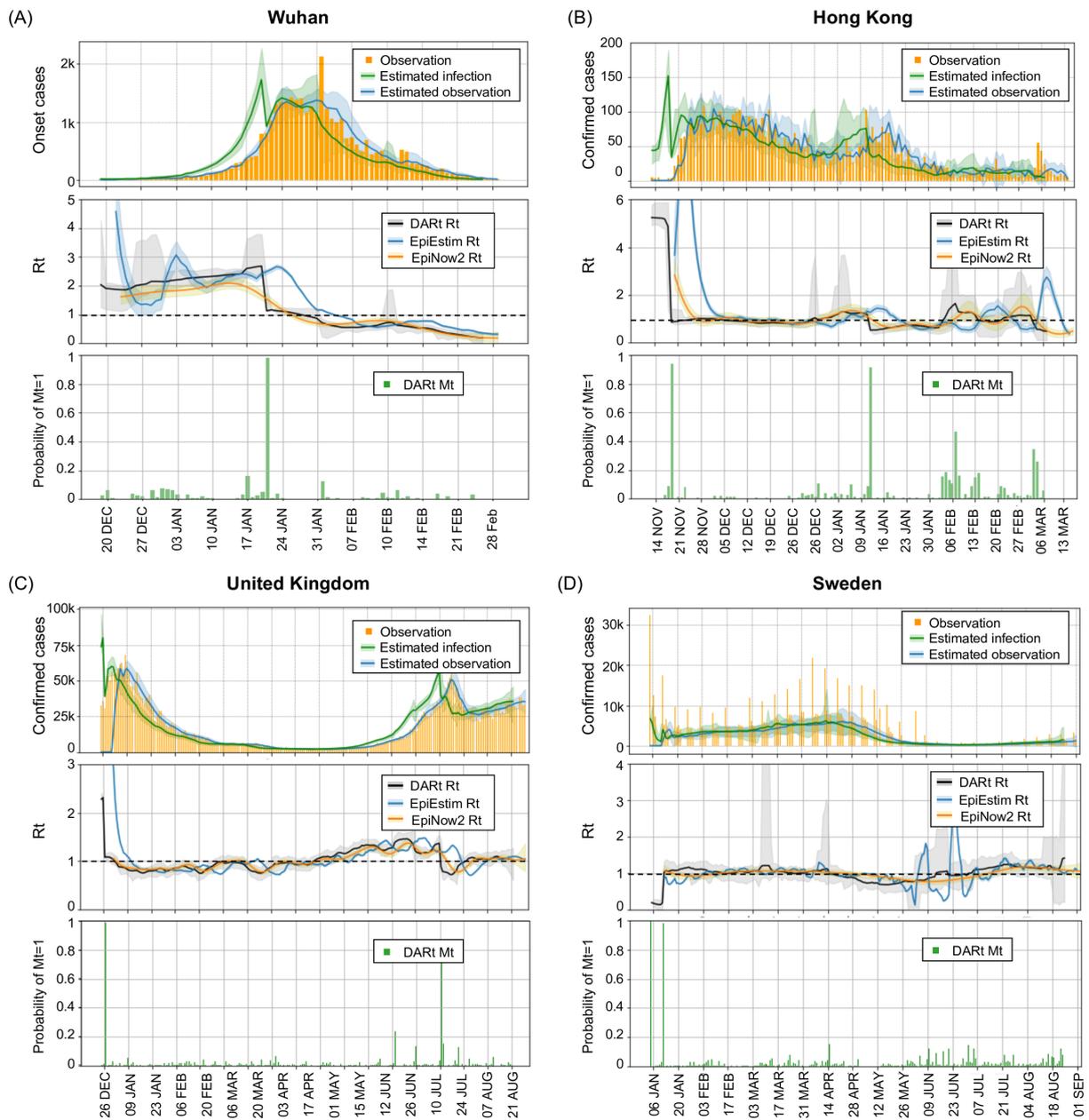

**Figure 4.** Epidemic dynamics in Wuhan **(A)**, Hong Kong **(B)**, United Kingdom **(C)** and Sweden **(D)**. The top row of each subplot shows the number of daily observations (in yellow), the estimated daily observations (in blue) and the estimated daily infections (in green). The middle row compares the DARt $R_t$ estimation (in black) with the EpiEstim results (in blue) and EpiNow2 (in yellow). The distributions of all estimated $R_t$ are with 95% CrI. The bottom row shows the probabilities of having abrupted changes ($M_t$=1) (in green bars).

**Epidemic dynamics in these four regions.** The inference results for $R_t$ and reconstructed observations for these four countries are shown in Figure 4. For Wuhan, the observation data are the number of onset cases compiled retrospectively from epidemic surveys, while for Hong Kong, UK and Sweden, the observation data are the number of reported confirmed cases. Notably, we use the onset-to-confirmed delay distribution from [33] together with the distribution of incubation time proposed in [3] to approximate the observation delay. As the ground-truth $R_t$ is not available, we validate the results by checking whether the estimated distributions of observation well cover the observation curve of $C_t$. As shown in the top panel of each subplot in Figure 4, the CrIs of estimated $C_t$ distributions (in blue) covered most parts of the original observations (in yellow), confirming the reliability of our $R_t$ estimation.

Figure 4 (A) shows the results of testing using Wuhan's onset data [1]. We observe that there was a sharp decrease in Wuhan's $R_t$ after 21$^{st}$ of Jan 2020, which is also illustrated by $M_t$ as the probability of abrupt changes peaked at this time (in green bars). A strict lockdown intervention has been enforced in Wuhan since 23$^{rd}$ of January 2020. This sharp decrease in Wuhan's $R_t$ is likely to be the result of this intervention, indicating its impact. The small offset between the exact lockdown date and the time of sharp decrease might be due to noisy onset observations and approximated incubation time distribution. After lockdown, $R_t$ decreased smoothly, indicating that people's awareness of the disease and the precaution measures taken had made an impact. Since the beginning of February 2020, the value of $R_t$ remained below 1 for most of the time with the enforcement of quarantine policy and increases in hospital beds to accept all diagnosed patients. It is noted that the onset curve has a peak on 1$^{st}$ February 2020, due to a major correction in the reporting standard. Neither $R_t$ nor $j_t$ curve from our model were severely suffered from this fluctuation, showing the robustness of our model thanks to the smoothing mechanism. The results from Wuhan suggest that our switching mechanism can overcome the issue of averaging and automatically detect sharp changes in epidemiological dynamics. The results from DARt are also compared with results from EpiEstim and EpiNow2. EpiEstim generates results with significant local variations and delays, while EpiNow2 can derive a smooth $R_t$ curve with no obvious delays. However, the immediate impact of lockdown cannot be well detected by EpiNow2.

Figure 4 (B) shows the inferred results from Hong Kong that reported confirmed cases[34] during the most recent outbreak from November 2020 to March 2021. In Hong Kong, the number of infections remains low for the most time and the government has continuously imposed soft interventions. In the middle of November 2020, a newly imported case has triggered a new outbreak, resulting in a large $R_t$ value. However, the $R_t$ level returned to be around 1 very soon as the government has further tightened social distancing measures at that time. From late January 2021, Hong Kong started to implement mandatory lockdown in the restricted areas. Since then, the number of daily cases remains at low level. Compared with the results from EpiEstim and EpiNow2, the results from DARt have similar trend with the others. The delays in the results of EpiEstim are still significant. We also investigate the performance of DARt using different types of observations and present the results of $R_t$ estimation using onsets and confirmed cases in the supplementary document.

Figure 4 (C) shows the inference results from the United Kingdom's reported confirmed cases[35]. It is noted that the United Kingdom was one of the first countries in the world to authorise the emergency use of COVID-19 vaccines. Since early December 2020, the United Kingdom rolled out its COVID-19 mass vaccination programme. By mid-February 2021, the United Kingdom had successfully hit its target of 15 million first-dose COVID-19 vaccinations, encompassing the top four priority groups for vaccination. As of 22nd April 2021, the UK had reached its target of 33 million (63% adults) first-dose COVID-19 vaccinations and 11 million (21% adults) second-dose. After 3 months since the mass vaccination programme started, the number of infection cases continuously followed a downward trend. However, since further easements of COVID-19 restrictions in mid-May 2021, $R_t$ gradually increased. During Euro

2020 football match (from June 11 to July 11), the $R_t$ value remained above 1. An immediate decrease in $R_t$ occurred when Euro 2020 was finished and since then $R_t$ remained around 1. The results from DARt, EpiEstim and EpiNow2 are generally consistent. However, DARt has accurately detected and responded to the impact of the completion of Euro 2020 in mid-July. In addition to studying the whole country, we applied DARt to typical cities in England to investigate the local epidemic dynamics as well (please refer to section 4 of the supplementary document).

Figure 4 (D) shows the inferred epidemic dynamics in Sweden from the daily reported data[36]. We find that the daily reported cases in Sweden had shown dramatic local fluctuations that were likely to be caused by misreporting. The reported cases dropped to 0 on Saturdays, Sundays and Mondays. This kind of observation noises could induce unnecessary fluctuations to $R_t$ curves. Therefore, we used Sweden's data to further illustrate the robustness of our scheme in the presence of noise. The results suggested that the influence of such periodic fluctuations has been smoothed by DARt and EpiNow2 to provide a consistent $R_t$ curve, where results from EpiEstim have shown significant local fluctuations.

To summarise, DARt has been applied to four different regions for investigating the transmission dynamics of COVID-19 to demonstrate its real-world applicability and effectiveness. Consistent with the findings in the simulation study, DARt has shown its advantages in the following aspects: 1) Instantaneity -- DARt adopts a window-free sequential Bayesian inference approach to detect and indicate abrupt epidemic changes; 2) Robustness -- with Bayesian smoothing, the $R_t$ curve from DARt is stable at the presence of observation noise; 3) Temporal accuracy -- DARt performs a joint estimation of $R_t$ and $j_t$ by explicitly encoding the lag into observation kernels.

## Discussion and conclusion

In this paper, we have proposed a Bayesian data assimilation scheme for estimating the time-varying epidemiological parameters based on observations. To study a real-world application scenario, we focus on estimating $R_t$ and provide a state-of-the-art $R_t$ estimation tool, DARt, supporting a wide range of observations. In the DARt system, epidemic states can therefore be updated using newly observed data, following a data assimilation process in the framework of sequential Bayesian updating. For the model inference, a particle filtering/smoothing method is used to approximate the $R_t$ distribution in both forward and backward directions of time, ensuring the $R_t$ at each time step assimilates information from all time points. By taking the Bayesian approach, we have emphasised the uncertainty in $R_t$ estimation by accommodating observation uncertainty in likelihood mapping and introduced Bayesian smoothing to incorporate sufficient information from observations. Our method provides a smooth $R_t$ curve together with its posterior distribution. We have demonstrated that inferred $R_t$ curves can describe different observations accurately. Our work is not only important in revealing the epidemical dynamics but also useful in assessing the impact of interventions. The sequential inference mechanism of $R_t$ estimation takes into account the accuracy of time alignment and provides an abrupt change indicator. Different from approaches of directly incorporating interventions as co-factors into epidemic model[13],[37], our method offers a promising method for intervention assessment.

We have made some approximations to facilitate the implementation. First, the observation time and generation time distributions are truncated into fixed and identical lengths. Theoretically, these two distributions can be of any length, while most values are quite small in practice. In our state transition model, one variable of the latent state is a vectorised form of infection numbers over a period. The purpose of vectorisation is to facilitate implementation by making the transition process Markovian. The length of this vector variable is determined by the length of effective observation time and the generation time distributions. Truncating these two distributions to a limited length, by discarding small values, would facilitate vectorisation. Apart from truncation, we have assumed that these two distributions do not change during the prevalence of disease. However, as we have discussed previously[23], introducing interventions, such as an increased testing capacity, would affect the observation time. The distribution of generation time would also change as the virus is evolving. It is possible to extend our model by adding a time-varying observation function. For example, the testing capability and time-varying mortality rate could also be considered in the observation process.

Second, we approximate the variance of observation error empirically. Given that variance of observations is unknown and could change over time across different regions, the standard deviation of the Gaussian likelihood function is not set to a fixed value in our scheme. Instead, we estimate the region-specific time-varying observation variance from the observational data. Although the empirical estimation yielded reasonable results for the four regions and cities in the UK (see the supplementary document), it may generate some implausible results in some scenarios, for example, when the epidemic is growing or resurging explosively, leading to an overestimation of observation variance. An adaptive error variance inference should be made to tackle this issue.

The third approximation is implicit in the use of a particle filter to approximate the posterior distributions over model state variables – including $R_t$ – with a limited number of samples (i.e., particles). Particle filtering makes no assumption about the form of posterior distributions. On the contrary, the variational equivalent of particle filter, namely variational filtering[38] provides an analytical approximation to the posterior probability and can be regarded as limiting solutions to an idealised particle filter, with an infinite number of particles[39]. Considering the importance of both the mean value of $R_t$ and its estimation uncertainty is important for advising governments on policymaking, an analytical approximation is desirable to help properly quantify uncertainty is desirable.

Finally, change detection is approximated by the change indicator $M_t$, which is included as part of the latent state and inferred during particle filtering. This work opens an avenue to explore variational Bayesian inference for switching state models[40]. Crucially, variational procedures enable us to assess model evidence (a.k.a. marginal likelihood) and hence allow automatic model selection. Examples of Variational Bayes and model comparison to optimise the parameters and structure of epidemic models can be found in previous studies[41]. These variational procedures can be effectively applied to change detection.

In conclusion, our work provides a practical scheme for accurate and robust estimation of time-varying epidemiological parameters. It opens a new avenue to study epidemic dynamics within the Bayesian data assimilation framework. We provide an open-source $R_t$ estimation package as well as an associated Web service that may facilitate other people's research in computational epidemiology and the practical use for policy development and impact assessment.

## Methods

The proposed Bayesian data assimilation framework for estimating epidemiological parameters include three main components: 1) a **state transition model** - describing the evolution of the latent state; 2) an **observation function** – defining an observation process and describing the relationship between the latent state and observations; 3) a **sequential Bayesian engine**: estimating statistical reason time-varying model parameters with uncertainty by assimilating prior state information provided by the transition model and the newly available observation. In this section, we introduce a real-world application of the proposed data assimilation framework to estimate one of the key epidemiological parameters, $R_t$. The modelling epidemic dynamics is characterised by the renewal process, which is the foundation of our state transition model. We then describe the observation function, linking a sequence of infection numbers with the observation data. Next, we present a detailed state transition model and propose the sequential Bayesian update module.

1. Renewal process for modelling epidemic dynamics

Common $R_t$ estimation methods include compartment model-based methods (e.g., SIR and SEIR [42] ) and time-since-infection models based on renewal process[31]. Their relationships are discussed in supplementary document Section 1.1. Comparative studies have been conducted in [21] to show that EpiEstim, one of the renewal process-based methods, outperforms other methods in terms of accuracy and timeliness. Given the renewal process, the key transition equation derived from the process is:

$$j_t = R_t \sum_{k=1}^{T_w} w_k \, j_{t-k} \qquad (1)$$

where $j_t$ is the number of incident infection cases on day $t$, $T_w$ is the time span of the set $\{w_k\}$, and individual $w_k$ is the probability that the secondary infection case occurs $k$ days after the primary infection, describing the distribution of generation time[10]. The profile of $w_k$ is related to the biological characteristics of the virus and is generally assumed to be time-independent during the epidemic. Considering the simplicity and superior performance of applying the renewal process to model epidemic dynamics, our work adopts Equation (1) as the basic transition function for joint estimation of $R_t$ and $j_t$.

## 2. Observation process

In epidemiology, the daily infection number $j_t$ cannot be measured directly but is reflected in observations such as the case reports of onset, confirmed infections and deaths. There is an inevitable time delay between the real date of infection and the date reporting, due to the incubation time, report delay, etc. Taking account of this time delay, we model the observation process as a convolution function between kernel $\varphi$, and the infection number in $T_H$ most recent days.

$$C_t = \sum_{k=d}^{T_H} \varphi_k \, j_{t-k} \tag{2}$$

where $C_t$ is the observation data, and $\varphi_k$ is the probability that an individual infected is detected on day $k$. $T_H$ is the maximum dependency window. It is assumed that the past daily infections before this window do not affect the current observation $C_t$. Since there is a delay between observation and infection, we suppose the most recent infection that can be observed by $C_t$ is at the time $t - d$, where $d$ is a constant determined by the distribution of observation delay.

To accommodate various observation types (e.g., the number of daily reported cases, onsets, deaths and infected cases), DARt will choose the appropriate time delay distributions accordingly. For example, for the input of onsets, the infected-to-onsets time distribution is chosen to be the kernel in the observation function. For the input of daily reported cases, the infected-to-onset and the onset-to-report delays are used together as the kernel in the observation function. These delay distributions can be either directly obtained from literature or inferred from case reports that contain individual observation delays[3]. Detailed descriptions of the observation functions for different epidemic curves can be found in the supplementary document (Section 1.2).

## 3. Sequential Bayesian Inference

In Figure 5, we illustrate the Bayesian inference scheme of DARt. The Bayesian data assimilation has two phases: forward filtering and backward smoothing. The forward filtering uses the up-to-date prior from the state transition model and the likelihood determined by the latest observation to update the current latent state, by computing its posterior distribution following the Bayes rule. The backward smoothing works by looking back to refine the previous state estimation when more observations are accumulated to reduce the uncertainty of parameter estimation.

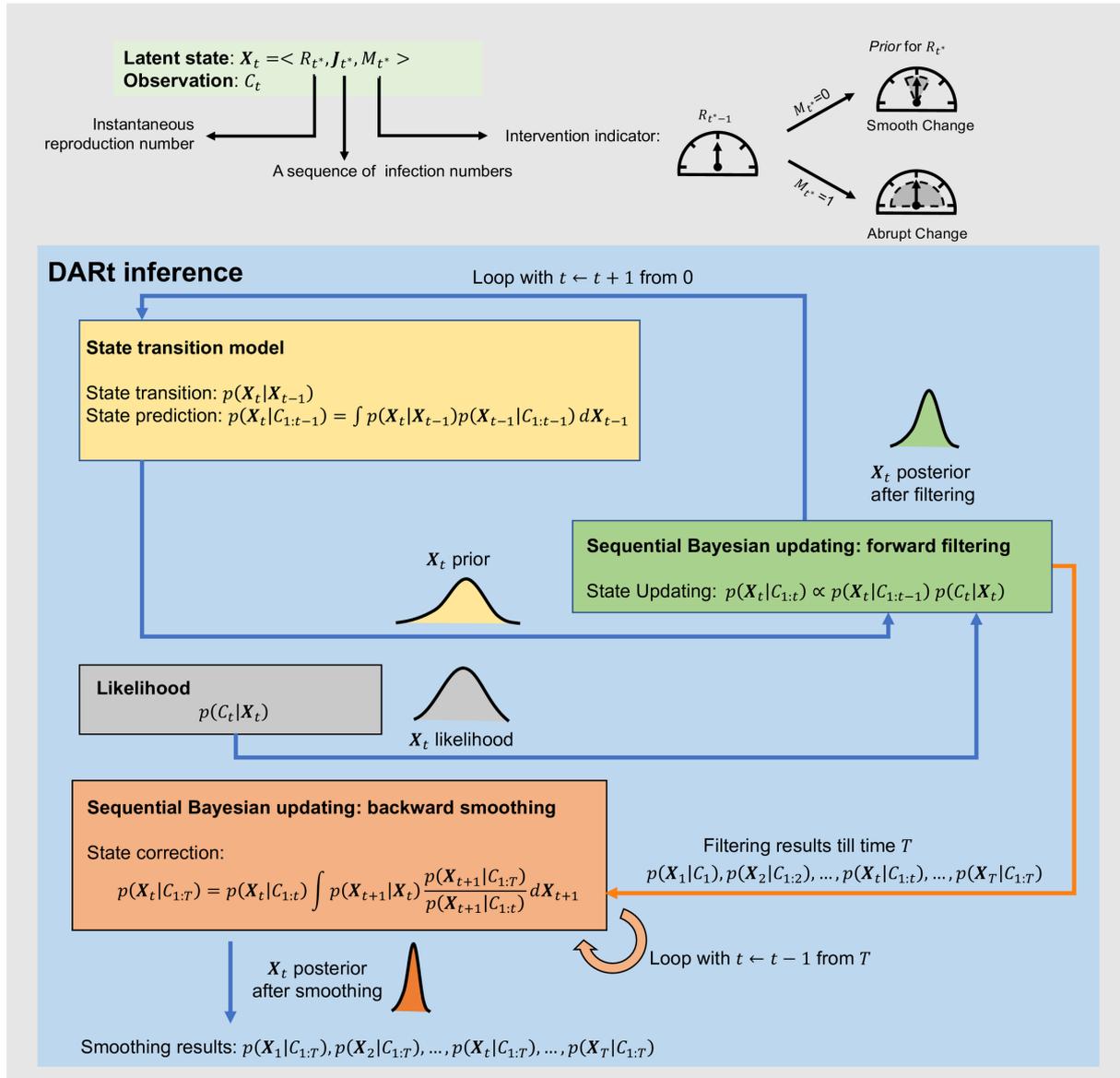

**Figure 5.** Three components of DARt inference model: state transition model, observation function and sequential Bayesian update module with two phases (forward filtering and backward smoothing). The latent state that can be observed in $C_t$ are defined as $X_t = <R_{t^*}, J_{t^*}, M_{t^*}>$ where $R_{t^*}$ is the instantaneous reproduction number, $M_{t^*}$ is a binary state variable indicating different evolution patterns of $R_{t^*}$, $J_{t^*} = [j_{t^*-T_\varphi+1}, j_{t^*-T_\varphi+2}, \ldots, j_{t^*}]$ is a vectorised form of infection numbers $j_t$, $t^*$ indicates the most recent infection that can be detected at time $t$ is from the time $t^*$ due to observation delay, and $T_\varphi$ is the length of the vector $J_{t^*}$ such that $C_t$ is only relevant to $J_{t^*}$ and $j_{t^*+1}$ only depends on $J_{t^*}$ via the renewal process.

▪ **State transition model**

In our model, indirectly observable variables $j_t$ and $R_t$ are included in the latent state. The state transition function for $R_t$ is commonly assumed to follow a Gaussian random walk [18] or constant within a sliding window as implemented in EpiEstim. Such a simplification cannot capture an abrupt change in $R_t$ under stringent intervention measures. To capture such abrupt changes, we introduce an auxiliary binary latent variable $M_t$ to indicate the switching dynamics of the epidemiological parameters under interventions without assuming a pre-defined evolution pattern (e.g., constant or exponential decay). $M_t = 0$ indicates a smooth evolution corresponding to minimal or consistent interventions; $M_t = 1$ indicates an abrupt change of corresponding to new interventions or outbreak. The smooth evolution is modelled as a Gaussian random walk while the abrupt change is captured through resetting the parameter memory by assuming a uniform probability distribution for the next time step of estimation. Doing so provides an automatic way of framing a new epidemic period that was manually done in [13]. The transition of $M_t$ is modelled as a discrete Markovian process with fixed transition probabilities controlling the sensitivity of change detection:

$$p(R_t|R_{t-1}, M_t) \sim \begin{cases} \mathcal{N}(R_{t-1}, \sigma_R^2) & M_t = 0 \quad \text{Mode I} \\ U[0, R_{t-1} + \Delta] & M_t = 1 \quad \text{Mode II} \end{cases} \quad (3)$$

where $\mathcal{N}(R_{t-1}, \sigma_R^2)$ is a Gaussian distribution with the mean value of $R_{t-1}$ and variance of $\sigma_R^2$, describing the random walk with the randomness controlled by $\sigma_R$. $U[0, R_{t-1} + \Delta]$ is a uniform distribution between 0 and $R_{t-1} + \Delta$ allowing sharp decrease while limiting the amount of increase. This is because we assume that $R_t$ can have a significant decrease when intervention is introduced but it is unlikely to increase dramatically as the characteristics of disease would not change instantly.

The transition of the change indicator $M_t$, is modelled as a discrete Markovian process with fixed transition probabilities:

$$p(M_t = 0|M_{t-1}) = p(M_t = 0) = \alpha \quad (4a)$$

$$p(M_t = 1|M_{t-1}) = p(M_t = 1) = 1 - \alpha \quad (4b)$$

where $\alpha$ is a value close to and lower than 1. The above function means that the value of $M_t$ is independent of $M_{t-1}$, while the probability of Mode II (i.e., $M_t = 1$) is quite small. This is because it is unlikely to have frequent abrupt changes in $R_t$.

For the incident infection $j_t$, the state transition can be modelled based on Equation (1) as $p(j_t|j_{t-1}, \dots, j_{t-T_w})$. To make the transition process Markovian, we vectorise the infection numbers as follows. Suppose the infection numbers $\{j_{t-k}\}_{k=d}^{T_H}$ that can be observed in $C_t$ are all included in $\boldsymbol{J}_{t^*} = [j_{t^*-T_\varphi+1}, j_{t^*-T_\varphi+2}, \dots, j_{t^*}]$, where $t^* = t - d$, and the length of this vector $T_\varphi$ is larger than or equal to $T_H - d + 1$. We also require $T_\varphi$ to be not smaller than $T_w$. Therefore, all the historical information needed to infer $j_{t^*}$ is available from $\boldsymbol{J}_{t^*-1}$, i.e., $\boldsymbol{J}_{t^*}$ only depends on $\boldsymbol{J}_{t^*-1}$ (i.e., being Markovian). The state transition process and observation process are illustrated in Figure 6.

The latent state in our model is then defined as $\boldsymbol{X}_t = <R_{t^*}, \boldsymbol{J}_{t^*}, M_{t^*}>$, which contribute to $C_t$ at time $t$. The state transition function of $\boldsymbol{J}_{t^*}$ is therefore Markovian:

$$p(\boldsymbol{J}_{t^*}|\boldsymbol{J}_{t^*-1}, R_{t^*}) = Poisson(j_{t^*}; R_{t^*} \sum_{k=1}^{T_w} w_k j_{t^*-k}) \prod_{m=1}^{T_\varphi - 1} \delta(\boldsymbol{J}_{t^*}^{(m)}, \boldsymbol{J}_{t^*-1}^{(m+1)}) \quad (5)$$

where $\boldsymbol{J}_{t^*}^{(m)}$ is the $m$-th component of the latent variable $\boldsymbol{J}_{t^*}$ and $\delta(x, y)$ is the Kronecker delta function (please refer to the supplementary document for more details). With Equation (3)-(5), the latent state transition function $p(\boldsymbol{X}_t|\boldsymbol{X}_{t-1})$ can be obtained as a Markov process:

$$p(\boldsymbol{X}_t|\boldsymbol{X}_{t-1}) = p(\boldsymbol{J}_{t^*}|\boldsymbol{J}_{t^*-1}, R_{t^*})p(R_{t^*}|R_{t^*-1}, M_{t^*})p(M_{t^*}|M_{t^*-1}) \quad (6)$$

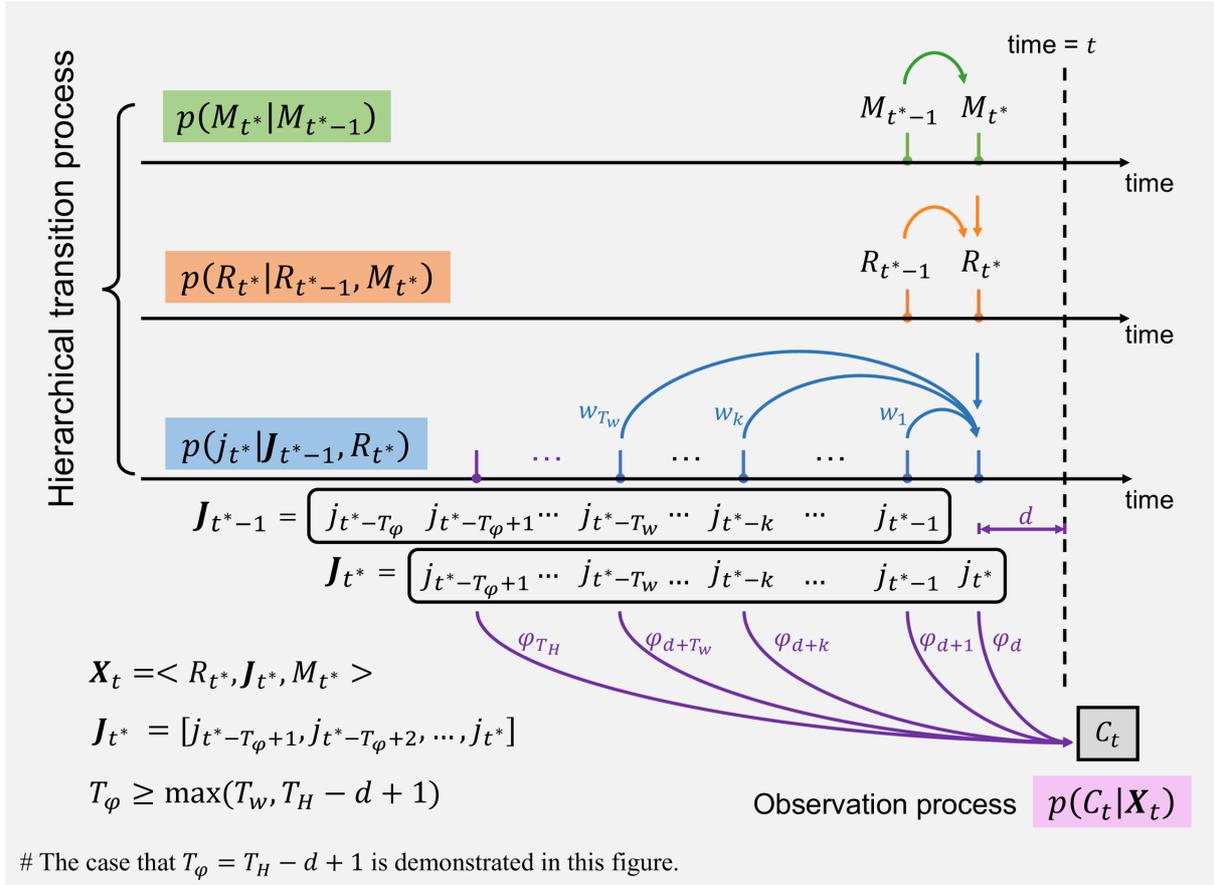

**Figure 6. Illustration of the hierarchical transition process and observation process.** The most recent infection that can be observed by $C_t$ is at the time $t^* = t - d$ where $d$ is a constant determined by the distribution of observation delay. Suppose $T_\varphi$ is the length of the vector $\boldsymbol{J}_{t^*} = [j_{t^*-T_\varphi+1}, j_{t^*-T_\varphi+2}, \ldots, j_{t^*}]$ such that $C_t$ is only relevant to $\boldsymbol{J}_{t^*}$ and $j_{t^*}$ only depends on $\boldsymbol{J}_{t^*-1}$ via the renewal process. Therefore, $T_\varphi \geq \max(T_w, T_H - d + 1)$. The case that $T_\varphi = T_H - d + 1$ is depicted in this figure.

▪ **Forward Filtering**

We formulate the inference of the latent state $\boldsymbol{X}_t = <R_{t^*}, \boldsymbol{J}_{t^*}, M_{t^*}>$ with the observations $C_t$ as within a data assimilation framework. A sequential Bayesian filtering approach is adopted to infer the time-varying latent state, which updates the posterior estimation using the latest observations following the Bayes rule. This approach differs from the fixed prior in the Bayesian inference of static parameters. This filtering mechanism computes the posterior distribution of the latent state by assimilating the forecast from the forward transition model with the information from the new epidemiological observations. For the implementation of this Bayesian updating process, we adopt a particle filter method[26] to efficiently approximate the posterior distribution through Sequential Monto Carlo (SMC) sampling. This eschews any fixed-form assumptions for the posterior – of the sort used in variational filtering and dynamic causal modelling[38].

Let us denote the observation history between time 1 and $t$ as $C_{1:t} = [C_1, C_2, ..., C_t]$. Given that previous estimation $p(X_{t-1}|C_{1:t-1})$ and new observation $C_t$, we would like to update the estimation of $X_t$, i.e., $p(X_t|C_{1:t})$ following the Bayes rule with the assumption that $C_{1:t}$ is conditionally independent of $C_{1:t-1}$ given $X_t$:

$$p(X_t|C_{1:t}) = \frac{p(C_t|X_t)p(X_t|C_{1:t-1})}{\int p(C_t|X_t)p(X_t|C_{1:t-1})dX_t} \qquad (7)$$

where $p(X_t|C_{1:t-1})$ is prior and $p(C_t|X_t)$ is the likelihood. The prior can be written in the marginalised format:

$$p(X_t|C_{1:t-1}) = \int p(X_t|X_{t-1})p(X_{t-1}|C_{1:t-1})\, dX_{t-1} \qquad (8)$$

where $X_t$ is assumed to be conditionally independent of $C_{1:t-1}$ given $X_{t-1}$, and the transition $p(X_t|X_{t-1})$ is defined in Equation (6) based on the underlying renewal process. The likelihood $p(C_t|X_t)$ can be calculated assuming the observation uncertainty follows a Gaussian distribution:

$$p(C_t|X_t) \sim \mathcal{N}(H(X_t), \sigma_C^2) \qquad (9)$$

where $H$ is the observation function with a kernel chosen according to the types of observations and $\sigma_C^2$ is the variance of observation error estimated empirically. To show the benefits of using this Gaussian likelihood function, we show the simulation results of using Poisson likelihood without considering the observation noise. Results can be found in Supplementary Figure 2, where the estimations fluctuate dramatically under noisy observation.

By substituting Equation (8) into Equation (7), we obtain the iterative update of $p(X_t|C_{1:t})$ given the transition $p(X_t|X_{t-1})$ and likelihood $p(C_t|X_t)$:

$$p(X_t|C_{1:t}) = \frac{p(C_t|X_t) \int p(X_t|X_{t-1})p(X_{t-1}|C_{1:t-1}) \, dX_{t-1}}{\iint p(C_t|X_t) \int p(X_t|X_{t-1})p(X_{t-1}|C_{1:t-1}) \, dX_{t-1} dX_t} \qquad (10)$$

- **Backward Smoothing**

The estimated result $p(X_t|C_{1:t})$ from aforementioned forward filtering only includes the past and present information flows, corresponding to the prior $p(X_t|C_{1:t-1})$ and likelihood $p(C_t|X_t)$, respectively. The filtering estimates would be accurate if all related infections are fully observed in $C_{1:t}$. However, this is certainly not the case due to observation delay. In order to reduce the uncertainty from forward filtering, we adopt the Bayesian backward smoothing technique, estimating the latent state at a time $t$ retrospectively, given all observations available till time $T$ ($T > t$). Compared with other parameter estimation methods [13], [16], Bayesian data assimilation takes the advantage of additional information to smooth inference results with reduced uncertainty caused by incomplete observations. More specifically, the smoothing mechanism can be described as: given a sequence of observations $C_{1:T}$ up to time $T$ and filtering results $p(X_t|C_{1:t})$, for all time $t < T$, the state estimates are smoothed as:

$$p(X_t|C_{1:T}) = p(X_t|C_{1:t}) \int p(X_{t+1}|X_t) \frac{p(X_{t+1}|C_{1:T})}{p(X_{t+1}|C_{1:t})} dX_{t+1} \qquad (11)$$

where $p(X_{t+1}|C_{1:T})$ is the smoothing results at time $t+1$ where $\int p(X_{t+1}|X_t) \frac{p(X_{t+1}|C_{1:T})}{p(X_{t+1}|C_{1:t})} dX_{t+1}$ is the smoothing factor. In this way, all the relevant observations are fully exploited to enable us to reduce the uncertainty of parameter estimation. Comparing with the sliding-window (i.e., averaging inference) approaches, our sequential Bayesian updating with backward smoothing mechanism features an instantaneous epidemiological parameter estimation and smoothing uncertainty through the utilisation of all available observations. More details can be found in the supplementary document.

## Data availability statement

We obtained daily onset or confirmed cases of four different regions (Wuhan, Hong Kong, Sweden, UK) from publicly available resources[1], [34]–[36]. For Wuhan, we adopted the daily number of onset patients from the retrospective study[1] (from the end of December 2020 to early March 2020). For UK data, we downloaded the daily report cases (cases by date reported) from the official UK Government website for data and insights on Coronavirus (COVID-19)[35] (from the start of January 2021 to the end of August 2021) accessed on 30th of August 2021. Data for UK Cities were also downloaded from the same resource[35] (from the start of January 2021 to the start of September 2021) accessed on 2nd of September 2021. For Sweden data, we downloaded the daily number of confirmed cases from the Our World in Data COVID-19 dataset [36] (from the middle of January 2021 to the start of September 2021) accessed on 2nd of September 2021. For Hong Kong, we downloaded the case reports from government website[34] (from the end of November 2020 to the end of March 2021), including descriptive details of individual confirmed case of COVID-19 infection in Hong Kong. For those asymptomatic patients whose onset date are unknown, we set their onset date as their reported date, and for those whose onset date is unclear, we simply removed and neglected these records. Only local cases and their related cases are considered, while imported cases and their related cases are excluded.

We are releasing DARt as open-source software for epidemic research and intervention policy design and monitoring. The source code of our method and our web service are publicly available online (https://github.com/Kerr93/DARt).

# Acknowledgement

Dr Simon Wang, at the Language Centre, HKBU, has helped improve the linguistic presentation of this manuscript.

# Supplementary Materials

# Bayesian data assimilation for estimating epidemic evolution: a COVID-19 study

Xian Yang, Shuo Wang, Yuting Xing, Ling Li, Richard Yi Da Xu,

Karl J. Friston and Yike Guo

## Table of Contents

# 1 Mathematical Models

## 1.1 Time-varying Renewal Process

Origin of Instantaneous Reproduction Number $R_t$. Both compartment models and time-since-infection models originate from the work of Kermack and McKendrick[1] and can be unified in the same mathematical framework[2]. Let us denote the numbers of susceptible and recovered individuals at calendar time $t$ by $S(t)$ and $U(t)$ (recovered individuals are not denoted as $R(t)$ in order to avoid confusion with reproduction number). Taking into account of different phases of the infection period, we denote the number of infected individuals with an infection-age $\tau$ by $i(t,\tau)$. Thus, the overall number of currently infected individuals at time $t$ is $I(t) = \int_0^t i(t,\tau)\,d\tau$ and the incident infection at time $t$ is $j(t) = i(t,0)$. Governing equations of the homogenous transmission[2] are as follows:

$$\frac{dS(t)}{dt} = -\lambda(t)S(t) \tag{S1}$$

$$\left(\frac{\partial}{\partial t} + \frac{\partial}{\partial \tau}\right) i(t,\tau) = -\gamma(\tau) i(t,\tau) \tag{S2}$$

$$\frac{dU(t)}{dt} = \int_0^t \gamma(\tau) i(t,\tau)\,d\tau \tag{S3}$$

$$i(t,0) = \lambda(t) S(t) \tag{S4}$$

where $\lambda(t)$ is the rate at which susceptible individuals get infected at time $t$. This is given by the infection rates per single infected individual $\beta(\tau)$ with an infection-age $\tau$ and the number of infected individuals $i(t,\tau)$ as:

$$\lambda(t) = \int_0^t \beta(\tau)\, i(t,\tau)\,d\tau \tag{S5}$$

Similarly, $\gamma(\tau)$ is defined as the recovery rate with at infection-age $\tau$. By simplifying Equation (S2) on the characteristic line ($t = \tau + c$), we have:

$$\frac{di(\tau + c, \tau)}{d\tau} = -\gamma(\tau) i(\tau + c, \tau) \tag{S6}$$

The solution to this ordinary differential equation is:

$$i(\tau + c, \tau) = i(c, 0) \mathcal{T}(\tau) \tag{S7}$$

where

$$\mathcal{T}(\tau) = exp\left(-\int_0^\tau \gamma(\sigma)\, d\sigma\right) \tag{S8}$$

Thus, we can link the infections with infection-age $\tau$ at time $t$ to the incident infection at time $t - \tau$:

$$i(t, \tau) = \mathcal{T}(\tau)\, i(t - \tau, 0) \tag{S9}$$

By substituting Equation (S5) and (S9) into Equation (S4), the incident infection $j(t)$ at time $t$ is

$$j(t) = \int_0^t S(t)\, \beta(\tau)\, \mathcal{T}(\tau)\, j(t - \tau) d\tau \tag{S10}$$

Then we have the infectiousness profile $\beta(t, \tau)$, representing the effectiveness rate at which an infectious individual with infection-age $\tau$ produces secondary cases at time $t$:

$$\beta(t, \tau) = S(t)\, \beta(\tau)\, \mathcal{T}(\tau) \tag{S11}$$

The corresponding instantaneous reproduction number $R(t)$ is derived from the integral of infectiousness profile $\beta(t, \tau)$:

$$R(t) = \int_0^\infty \beta(t, \tau) d\tau \tag{S12}$$

It is the average number of people that someone infected at time $t$ is expected to infect, if conditions remain unchanged (i.e. susceptible population, infectiousness rate, recovery rate). From the above derivation, we observe that the infectiousness profile $\beta(t,\tau)$ and corresponding instantaneous reproduction number $R_t$ are composed of three factors: $S(t)$, $\beta(\tau)$ and $\mathcal{T}(\tau)$. $S(t)$ represents the depletion of susceptible individuals: the decline of $S(t)$ will reduce the susceptible population size leading to possible herd immunity. $\beta(\tau)$ represents the infectiousness of individuals with infection-age $\tau$. This is related to biological (e.g. viral shedding) and behavioural (e.g. contact rates) factors. $\mathcal{T}(\tau)$ represents the recovery rate of individuals with infection-age $\tau$: faster recovery will result in shorter infectiousness period and smaller reproduction number. All three factors, $S(t)$, $\mathcal{T}(\tau)$ and $\beta(\tau)$ can be altered by the implementation of control measures along with time.

Decomposition of the Infectiousness Profile. $R_t$ is determined by the evolution of the infectiousness profile $\beta(t,\tau)$ according to the Equation (S12). Further, the infectiousness profile $\beta(t,\tau)$ can be rewritten as:

$$\beta(t,\tau) = R(t)\, w(t,\tau) \qquad (S13)$$

where $w(t,\tau) = \beta(t,\tau)/\int \beta(t,\tau)d\tau$ is called the distribution of generation time, representing the probability distribution of infection events as a function of infection-age $\tau$. That is, the distribution of time interval between the primary infection and subsequent secondary infection. In principle, the distribution of generation time is time-varying due to the three factors in Equation (S11), which increases the complexity of parametric modelling. Most existing studies assume a time-invariant generation time distribution (i.e. $w(t,\tau) = w(\tau)$) while the introduction of control measures results in the change of $R(t)$. Under this assumption, Equation (S10) can be rewritten as:

$$j(t) = R(t) \int_0^t w(\tau) \, j(t-\tau) \, d\tau \tag{S14}$$

This is the core formula for $R_t$ estimation from the infection data. That is,

$$R(t) = j(t) / \int_0^t w(\tau) \, j(t-\tau) \, d\tau \tag{S15}$$

Or, we can have the corresponding discretised version:

$$R_t = \frac{j_t}{\sum_{k=1}^{T_w} w_k j_{t-k}} \tag{S16}$$

where $T_w$ is the time span of the set $\{w_k\}$. This decomposition of infectiousness profile into $R_t$ and time-invariant generation time distribution $w_k$ is one of the fundamental formulae for $R_t$ estimation in the existing literature (e.g. the well-known package 'EpiEstim'[3] and this paper).

## 1.2 Observations of the Transmission Dynamics

Formulation of the Observation Function. The infection number $j_t$ is the ideal data source for $R_t$ estimation according to Equation (S16). However, it is impossible to obtain the exact number of real-time infections through intensive screening. Instead, the infections are usually observed from the statistical reports of related events (i.e. epidemic curves) such as the daily report of confirmed cases, onset cases and death number. There is an inevitable delay between the occurrence of infecting events and the events being reported. That means these epidemic curves do not reflect the current incidence of infection $j_t$. We clarify this by formulating the observation function of the transmission dynamics. In the framework of data assimilation, $j_t$ is the state variable of the dynamic epidemic system and its update is driven by the parameter $R_t$ as described by Equation (S16). The aggregated reports $C_t$ (e.g., daily confirmed cases, deaths) are the observing results of the state variable through an observation function $H$:

$$C_t = H(j_t) \tag{S17}$$

where $H$ is the observation function and $C_t$ is the observation result.

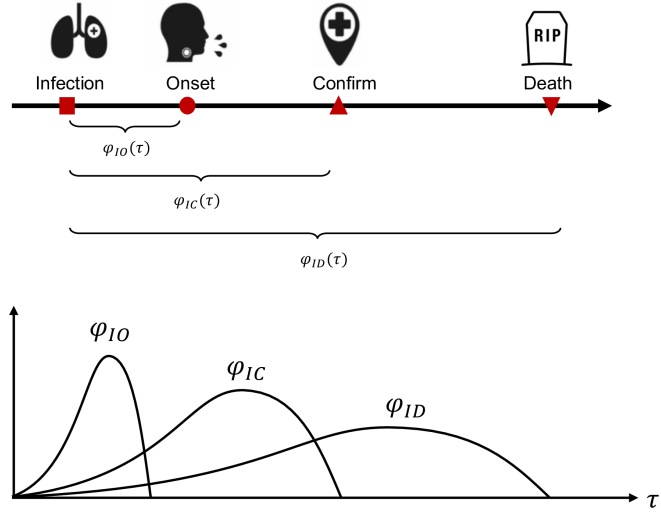

**Supplementary Figure 1**. Illustrations of three types of observations and corresponding distributions of delay from the real infection date and observation.

Observation Functions for Various Reports. The format of the observation function $H$ depends on the type of reported data being used. In general, H is a convolutional operation summing up the portion of infected cases weighted by the distribution of time delay between being infected and being reported. Confirmed reports $C_t^C$, onset cases $C_t^O$ and death reports $C_t^D$ are the three most used reported data for $R_t$ estimation[4]–[6] (illustrated in Supplementary Figure 1).

**A. Onset Cases Reports.** The reports of onset cases are usually compiled retrospectively from epidemic surveys of confirmed cases, which can be represented as:

$$C_t^O = \sum_{k=d_O}^{T_O} j_{t-k} \varphi_k^{IO} = \varphi^{IO} \otimes j_t \tag{S18}$$

where $\varphi_k^{IO}$ is the probability that the symptom onset occurs $k$ days after the initial infection date for a reported case, and $d_O$ indicates the $C_t^O$ can only cover information of infections at least $d_O$ days before $t$. The value of $d_O$ is determined by $\varphi^{IO}$. The distribution $\varphi^{IO}$ is determined by the biological factors of the virus and has been investigated in the previous reports[7], which is considered time-independent. We use the symbol ⊗ to denote the convolution operation.

**B. Confirmed Cases Reports.** The epidemic curve of daily confirmed cases $C_t^C$ is observed from

$$C_t^C = \sum_{k=d_C}^{T_C} j_{t-k}\varphi_k^{IC} = \varphi^{IC} \otimes j_t \quad (S19)$$

where $\varphi_k^{IC}$ is the probability that a confirmed case is reported $k$ days after the initial infection date, and $d_C$ has a similar definition with $d_O$. The distribution $\varphi^{IC}$ includes two parts: the time between infection to symptom onset $\varphi^{IO}$, and the time between symptom onset to reported confirmation $\varphi^{OC}$:

$$\varphi^{IC} = \varphi^{IO} \otimes \varphi^{OC} \quad (S20)$$

The former part $\varphi^{IO}$ is usually similar across regions while the latter time delay $\varphi^{OC}$ varies a lot due to test policies and screening capabilities.

**C. Death Reports.** The epidemic curve of death $C_t^D$ is observed from

$$C_t^D = \rho_D \sum_{k=d_D}^{T_D} j_{t-k}\varphi_k^{ID} = \rho_D \varphi^{ID} \otimes j_t \quad (S21)$$

where $\rho_D$ is the observed mortality rate of infected cases, $\varphi_k^{ID}$ is the probability that a confirmed case is reported dead $k$ days after the initial infection date, and $d_D$ has a similar definition with $d_O$. $\rho_D$ and $\varphi^{ID}$ vary across different countries and periods due to capacities of treatment[5].

## 2 Model Inference

### 2.1 Problem Formulation

The time-varying renewal process can be formulated through the framework of state space hidden Markov models. The instantaneous reproduction number $R_t$ and daily incident infections $j_t$ are the two latent variables of the state space models, whose dependence is described by Equation (S16). Consider two evolution modes of $R_t$: emerging smooth changes when interventions are being steadily introduced/relaxed, and undergoing an abrupt change due to intensive interventions (e.g., lockdown). We introduce another latent variable $M_t$ to automate the switch between these two modes, which will be discussed in detail in the next section. The observations $C_t$ are the observed results of $j_t$ through Equation (S18), (S19) and (S21). We are interested in inferring the evolution of $R_t$ (along with $j_t$ and $M_t$) upon the real-time update of observations $C_t$.

## 2.2 Inference Aims

As revealed in Equation (S14) and Equation (S17), the observations experience time delay with respect to the update of the latent state, due to the lagging and averaging effects of convolution in Equation (S18)-(S21). Thus, the changes of $R_t$ cannot be reflected in time, due to the incubation time and observation delay. In other words, accurate estimation of $R_t$ at time $t$ relies on future observations, which imposes the challenges of timely estimation. Therefore, we focus on two inference aims:

1. Given the latest observation, how to give a near real-time estimate of $R_t$ and – equally if not more important – how to assess the uncertainty of the results?

2. Upon update of the real-time observations, how to modify estimations at all previous time steps and assess the uncertainties to make them more accurate taking into account the new information?

These two aims correspond to the two fundamental problems in Bayesian updating, namely the **filtering** and **smoothing** problems to be discussed in the next section.

## 2.3 Bayesian Updating Scheme

$X_t = <R_{t^*}, J_{t^*}, M_{t^*}>$ is defined as the latent state observed by $C_t$ at time $t$. Since there is a delay between observation and infection, we suppose the most recent infection that can be observed by $C_t$ is at the time $t^* = t - d$, where $d$ is a constant determined by the distribution of observation delay. Suppose $T_\varphi$ is the length of the vector $\boldsymbol{J}_{t^*} = [j_{t^*-T_\varphi+1}, j_{t^*-T_\varphi+2}, \ldots, j_{t^*}]$ such that $C_t$ is only relevant to $\boldsymbol{J}_{t^*}$ via Equation (S18)-(S21) and $j_{t^*}$ only depends on $\boldsymbol{J}_{t^*-1}$ via the renewal process. We formulate the estimate of the latent state $\boldsymbol{X}_t$ from the observed reports $C_t$ as a data assimilation problem. A sequential Bayesian approach is adopted to infer time-varying latent state, which are composed of two phases: forward filtering and backward smoothing.

**Forward filtering:** A sequential Bayesian updating approach is employed to infer the latest latent state from the real-time observations. Let us denote the observation history between time 1 and $t$ as $C_{1:t} = [C_1, C_2, \ldots, C_t]$. Given that previous estimation $p(\boldsymbol{X}_{t-1}|C_{1:t-1})$ and new observation $C_t$, we would like to update the estimation of $\boldsymbol{X}_t$, i.e., $p(\boldsymbol{X}_t|C_{1:t})$ following the Bayes rule:

$$p(\boldsymbol{X}_t|C_{1:t}) = \frac{p(C_t|\boldsymbol{X}_t)p(\boldsymbol{X}_t|C_{1:t-1})}{\int p(C_t|\boldsymbol{X}_t)p(\boldsymbol{X}_t|C_{1:t-1})\,d\boldsymbol{X}_t} \qquad (S22)$$

where $p(\boldsymbol{X}_t|C_{1:t-1})$ is *prior* and $p(C_t|\boldsymbol{X}_t)$ is *likelihood*. The *prior* can be written in the marginalised format:

$$p(\boldsymbol{X}_t|C_{1:t-1}) = \int p(\boldsymbol{X}_t|\boldsymbol{X}_{t-1})p(\boldsymbol{X}_{t-1}|C_{1:t-1})\,d\boldsymbol{X}_{t-1} \qquad (S23)$$

which utilised the Markovian properties. By substituting Equation (S23) to Equation (S22), we obtain the iterative update of $p(\boldsymbol{X}_t|C_{1:t})$ given the transition $p(\boldsymbol{X}_t|\boldsymbol{X}_{t-1})$ and *likelihood* $p(C_t|\boldsymbol{X}_t)$:

$$p(\boldsymbol{X}_t|C_{1:t}) = \frac{p(C_t|\boldsymbol{X}_t)\int p(\boldsymbol{X}_t|\boldsymbol{X}_{t-1})p(\boldsymbol{X}_{t-1}|C_{1:t-1})\,d\boldsymbol{X}_{t-1}}{\iint p(C_t|\boldsymbol{X}_t)\int p(\boldsymbol{X}_t|\boldsymbol{X}_{t-1})p(\boldsymbol{X}_{t-1}|C_{1:t-1})\,d\boldsymbol{X}_{t-1}d\boldsymbol{X}_t} \qquad (S24)$$

The *likelihood* can be calculated assuming an observation is with Gaussian variance:

$$p(C_t|\boldsymbol{X}_t) \sim \mathcal{N}(H(\boldsymbol{X}_t), \sigma_C^2) \qquad (S25)$$

where $H$ is chosen accordingly to the types of reports and $\sigma_c^2$ is the variance of observation error that can be approximated empirically (detailed settings can be found in Supplementary Section 4). As the likelihood function has explicitly considered observation noise, the function is more robust to noise compared to Poisson likelihood (used in EpiEstim). The results of using the Poisson likelihood for the same synthesised data as depicted in Figure 3 are shown in Supplementary Figure 2, showing the benefits of considering observation noise in the likelihood.

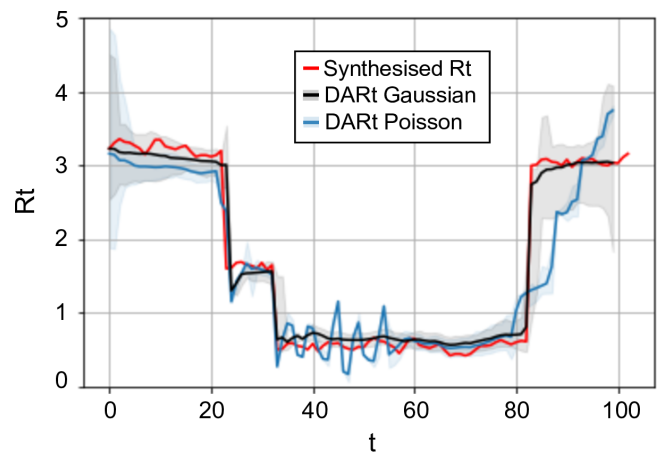

**Supplementary Figure 2.** Comparison between the simulation results using Poisson likelihood and Gaussian likelihood in DARt (both with 95% CrI).

Then we present the $p(X_t|X_{t-1})$, i.e., the transition of the latent state $X_t =< R_{t^*}, J_{t^*}, M_{t^*} >$ in details through analysing the evolution patterns of $R_{t^*}$. When $R_{t^*}$ is evolving with smooth changes, we use a Gaussian random walk to model this pattern, named Mode I corresponding to $M_{t^*} = 0$. Under this mode, it is expected that $R_{t^*}$ is similar to the previous time $R_{t^*-1}$. In contrast, the evolution of $R_{t^*}$ can be altered significantly when intensive measures are induced. For example, $R_{t^*}$ may experience an abrupt decrease due to the lockdown policy on time $t^*$. Under this circumstance, the epidemic history does not provide much information about the latest $R_{t^*}$, where we name it as Mode II corresponding to $M_{t^*} = 1$.

Formally, the evolution of $R_{t^*}$ is described by the switching dynamics conditioned on $M_{t^*}$:

$$p(R_{t^*}|R_{t^*-1}) \sim \begin{cases} \mathcal{N}(R_{t^*-1}, \sigma_R^2) & M_{t^*} = 0 \quad \text{Mode I} \\ U[0, R_{t^*-1} + \Delta] & M_{t^*} = 1 \quad \text{Model II} \end{cases} \quad (S26)$$

where $\mathcal{N}(R_{t^*-1}, \sigma_R^2)$ is a Gaussian distribution with the mean value of $R_{t^*-1}$ and variance of $\sigma_R^2$, describing the random walk with the randomness controlled by $\sigma_R$. $U[0, R_{t^*-1} + \Delta]$ is a uniform distribution between 0 and $R_{t^*-1} + \Delta$ allowing abrupt decrease while limiting the amount of increase. This is because we assume that $R_{t^*}$ can undergo a big decrease when intervention is introduced but it is unlikely to have a dramatic increase in one day as the characteristics of disease would not change instantly.

In our model, we assume a discrete Markovian chain process for $M_{t^*}$ with the transition probabilities listed in Supplementary Table 1, meaning that the probability of having an abrupt change in $R_{t^*}$ is low. This assumption is realistic as most of the time the $R_{t^*}$ curve is undergoing smooth change.

**Supplementary Table 1.** Transition probabilities of $M_{t^*}$.

|  | $M_{t^*} = 0$ | $M_{t^*} = 1$ |
| --- | --- | --- |
| $M_{t^*-1} = 0$ | 0.95 | 0.05 |
| $M_{t^*-1} = 1$ | 0.95 | 0.05 |

Finally, we use the renewal process to provide transition of $J_{t^*+1}$:

$$p(J_{t^*}|J_{t^*-1}, R_{t^*}) = Poisson(j_{t^*}; R_{t^*} \sum_{k=1}^{T_w} w_k j_{t^*-k}) \prod_{m=1}^{T_\varphi - 1} \delta(J_{t^*}^{(m)}, J_{t^*-1}^{(m+1)}) \quad (S27)$$

where $J_{t^*}^{(m)}$ is the $m$-th component of the latent variable $J_{t^*}$ and $\delta(x, y)$ is the Kronecker delta function. $j_{t^*}$ is assumed to be drawn from a Poisson distribution with the mean equal to the prediction from the renewal process using $R_{t^*}$ and $J_{t^*-1}$. The overlaps between $J_{t^*-1}$ and $J_{t^*}$ are $\{J_{t^*}^{(m)}\}_{m=1}^{T_\varphi - 1} = \{J_{t^*-1}^{(m+1)}\}_{m=1}^{T_\varphi - 1} = [j_{t^* - T_\varphi + 1}, \ldots, j_{t^*-1}]$, whose distributions are assumed to be consistent. By substituting Equation (S26) and (S27) and Supplementary Table 1 into Equation (S24), we have realised the sequential Bayesian update of the latent state $X_t = <R_{t^*}, J_{t^*}, M_{t^*}>$ for filtering.

**Backward smoothing:** To answer the second question on how to update previous estimations when more subsequent observations are available, we formulate it as a smoothing problem in the Bayesian updating framework. Based on the filtering results of $p(X_t|C_{1:t})$, we can further achieve the smoothing results $p(X_t|C_{1:T})$, where $T$ is the time index of the last observation. To integrate the information from the subsequent observations, we use the backward pass method. First, the joint distribution $p(X_1, \ldots, X_T|C_{1:T})$ is decomposed as:

$$p(X_1, \ldots, X_T|C_{1:T}) = p(X_T|C_{1:T}) \prod_{t=1}^{T} p(X_t|X_{t+1}, C_{1:T})$$

$$= p(X_T|C_{1:T}) \prod_{t=1}^{T} p(X_t|X_{t+1}, C_{1:t}) \quad (S28)$$

where

$$p(X_t|X_{t+1}, C_{1:t}) = \frac{p(X_{t+1}|X_t) p(X_t|C_{1:t})}{p(X_{t+1}|C_{1:t})}. \quad (S29)$$

Then by integrating out $X_1, \ldots X_{t-1}, X_{t+1}, X_T$ in Equation (S28)

$$p(X_t|C_{1:T}) = p(X_t|C_{1:t}) \int p(X_{t+1}|X_t) \frac{p(X_{t+1}|C_{1:T})}{p(X_{t+1}|C_{1:t})} dX_{t+1} \quad (S30)$$

which provides the iterative calculation of $p(X_t|C_{1:T})$ from time $T$ backwards to time $t$.

Our model includes both $j_t$ and $R_t$ into the latent state and jointly estimates them together. The Supplementary Figure 3 compares $j_t$ from our method with $j_t$ by using the back calculation function from EpiNow2[1] (for the synthesised data mentioned in the main manuscript). We can see that our method returns a $j_t$ curve with CrI that is closer to the simulated infection curve.

---

[1] https://epiforecasts.io/EpiNow2/reference/backcalc_opts.html

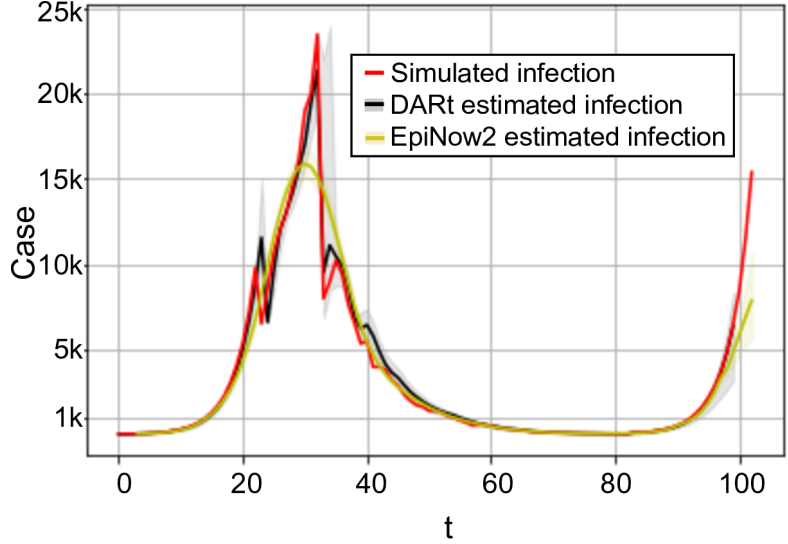

**Supplementary Figure 3.** Comparison between the estimated daily infection. The estimated infection by DARt is drawn in black with 95% CrI. The ground-truth simulated infection is in red and the back calculated infection is in yellow.

## 2.4 Particle Methods

The integrals in the filtering problem (Equation (S24)) and the smoothing problem (Equation (S30)) are intractable, thus we introduce a Sequential Monte Carlo (SMC) method called 'particle filter' to infer the latent state[8].

In Monte Carlo method, the continuous distribution of a random variable $X \sim \pi(x)$ is approximated by $N$ independent samples with importance weights:

$$\pi(x) \approx \sum_{i=1}^{N} W^i \delta_{X^i}(x) \tag{S31}$$

where $\pi(x)$ is an arbitrary probability distribution and $N$ independent samples $X^i \sim \pi(x)$ are drawn from the distribution with the normalized importance weight $W^i$. $\delta_{X^i}(x)$ denotes the Dirac delta mass located at the i-th sample $X^i$. These discrete samples are also called 'particles' in particle method, whose locations and weights are used to approximate the intractable integral. If $x$ is a time-dependent state variable, we can update the samples to approximate the distribution through the Sequential Importance Sampling (SIS) technique[9]. The locations and weights of the particles representing the target distribution are iteratively updated considering the new observations. For the filtering problem, we can set $p(X_{1:t}|C_{1:t})$ as the target distribution and use $N$ particles $\{X_t^1, X_t^2, ..., X_t^N\}$ with importance weight $\{W_t^1, W_t^2, ..., W_t^N\}$ at time $t$. The SIS technique includes two steps: First, new positions of the particles $X_t$ at time $t$ are proposed according to a proposal function $q(X_t|X_{1:t-1})$ which can be the transition probability $p(X_t|X_{1:t-1})$. Next, the importance weight $\omega(X_{1:t})$ of the proposed particles are calculated according to iteration in Equation (S22):

$$\omega(X_{1:t}) = \frac{p(X_{1:t}|C_{1:t})}{q(X_{1:t})} = \frac{p(X_{1:t-1}|C_{1:t-1})}{q(X_{1:t-1})} \frac{p(X_t|X_{1:t-1})p(C_t|X_t)}{q(X_t|X_{1:t-1})}$$

$$= \omega(X_{1:t-1}) \frac{p(X_t|X_{1:t-1})p(C_t|X_t)}{q(X_t|X_{1:t-1})} = \omega(X_{1:t-1})p(C_t|X_t) \quad (S32)$$

Therefore, the filtering results of $p(X_{1:t}|C_{1:t})$ can be numerically approximated by the evolving particles and their importance weights. Similarly, the smoothing procedure of Equation (S30) can be approximated by the particles.

## 3 DARt Application to HK

Supplementary Figure 4 shows the results of $R_t$ estimation using onsets and confirmed cases as observations in Hong Kong to estimate $R_t$, respectively. We choose Hong Kong for illustration purpose since both onset and reported confirmed cases are publicly accessible. The results from Supplementary Figure 4 reflect that, although the exact values of $R_t$ at some time points would be different due to difference in observations, the overall trends are largely consistent. With a proper observation kernel according to the observation types (e.g., onset, confirmed cases), DARt can make generally consistent estimation from a wide range of observation types.

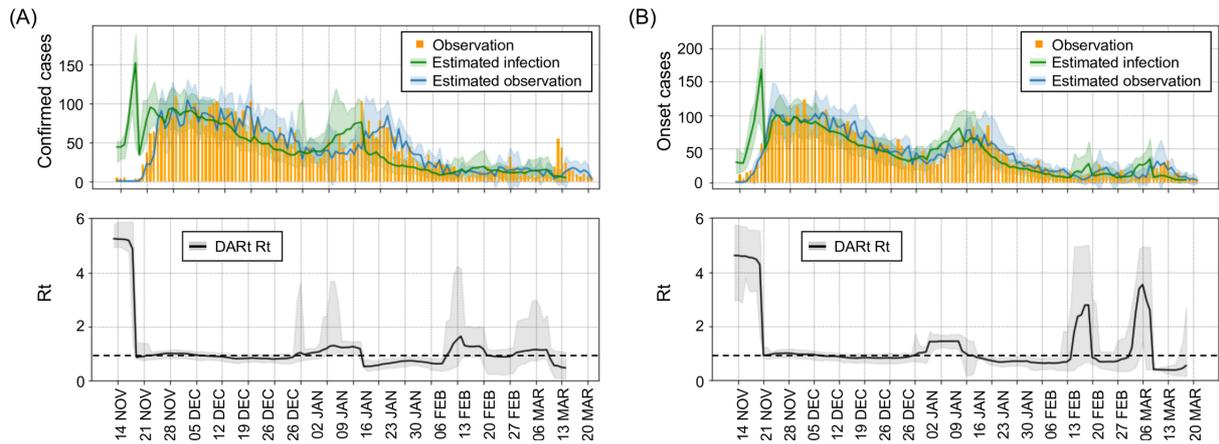

**Supplementary Figure 4.** Comparison of estimated $R_t$ curves of Hong Kong using different observations. Subplot A) shows $R_t$ estimations (in black) from confirmed cases (in yellow). Subplot B) shows $R_t$ estimations (in black) from daily onset (in yellow).

# 4  DARt Application to UK Cities

We applied DARt to monitor seven cities in England as shown in Supplementary Figure 5, reflecting that the country-wide $R_t$ curve shown in Figure 5 (C) cannot be used to represent the epidemic dynamics across different local areas. With the application of DARt, we can examine the impacts of events, especially the easing of COVID restrictions that happened on 17/05/2021. Different cities responded to the easements differently as reflected in their individual $R_t$ curves given their different micro social structures. We can find that most cities have experienced an increase in $R_t$ right after the easement of restrictions. In particular, sharp increases can be found in Leeds, Liverpool and Sheffield in late May. We can also find that the $R_t$ levels for all cities were back to 1 in July and August. For London, which is one of the Euro Cup 2020 host cities, had an elevated $R_t$ level during the Euro cup 2020. An obvious decrease in $R_t$, as indicated by $M_t$, happened around middle of June after the final of Euro Cup 2020. This observation implies big sport event would increase the $R_t$ values.

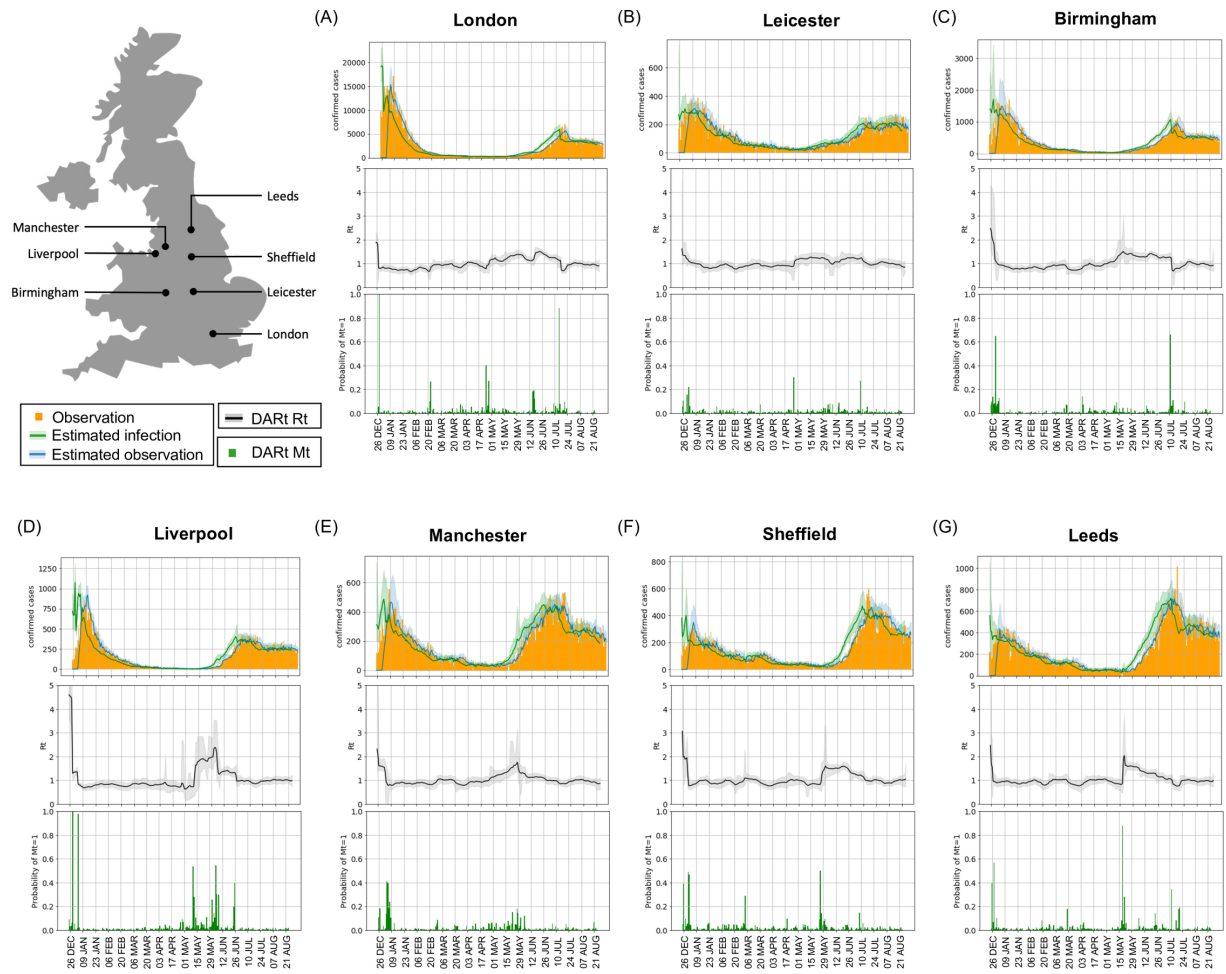

**Supplementary Figure 5.** Epidemic dynamics in London, Leicester, Birmingham, Liverpool, Manchester, Sheffield, and Leeds. The top row of each subplot shows the number of daily observations (in yellow), the estimated daily observations (in blue) and the estimated daily infections (in green). The middle row shows the DARt results of $R_t$ curve with 95% CrI (in black), while the probability of having abrupt changes is shown in the bottom row (i.e., $M_t = 1$) (in green).

## 5 Experimental setting

In our experiments, the generation time and observation delay distributions are adopted from the previous reports [6], [10]. To truncate these distributions into a fixed length, we discard the time points with the kernel values smaller than 0.1 resulting in the length of $J_t$ as 7. The initial guess of $R_t$ at $t = 0$ is set to be uniformly distributed from 1 to 5. We set $\sigma = 0.1$ for getting smooth $R_t$ change in Model I. In Model II, we set $\Delta = 0.5$ for all regions. To implement the particle filter, the number of particles is set to 200 for approximating distributions.

The variance of observation error $\sigma_C^2$ is estimated empirically. We first calculate the 7-day moving average observations. By subtracting the moving average from the observation, we obtain a difference curve, approximating random observation fluctuations. The next step is to perform the 7-day moving average calculation again on the squared value of the difference curve, where the resulted curve is regarded as the observation error variance. Finally, we use a Gaussian distribution as the likelihood function (Equation (S25)), where its variance is approximated by the observation error variance curve.

## 6 Sensitivity Analysis

### 6.1 Different levels of observation noise

To investigate the robustness of different comparative methods, Figure 6 shows the $R_t$ estimation results from DARt, EpiEstim and EpiNow2 under different levels of observation noise. We can find that with greater noises, all comparative methods would experience some estimation errors. Particularly for EpiEstim, local fluctuations became obvious on the first few days. For DARt, although high observation noises would influence the $M_t$ estimation results, the results are still informative to indicate abrupt changes.

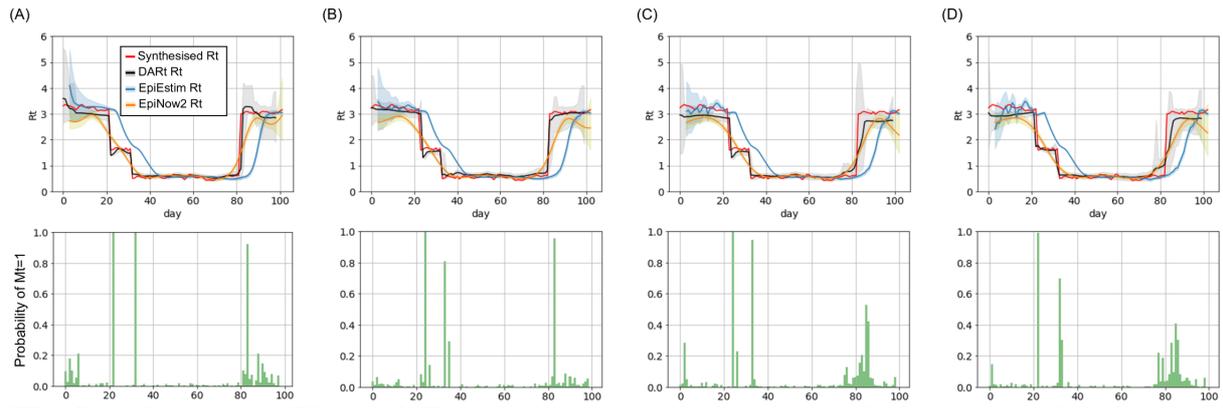

**The Supplementary Figure 6.** The $R_t$ estimation results under different levels of observation noise: A) N=0, B) N=1, C) N=2 and D) N=3, where the added Gaussian noise has the standard deviation equal to $N$ times of the unperturbed observation.

### 6.2 Different values of truncation threshold

Here, we would like to conduct experiments to investigate how different settings of the truncation threshold would influence the $R_t$ estimation results. The supplementary Figure 7(a), (b) and (c) present the $R_t$ results with the truncation threshold set to 0.1, 0.05 and 0.01, respectively. Other experimental settings are same as that in the main manuscript with the absence of observation noise. We can find that the results from these three subplots are quite similar implying that setting the threshold to be 0.1 as what we did throughout the paper would not influence the estimation significantly.

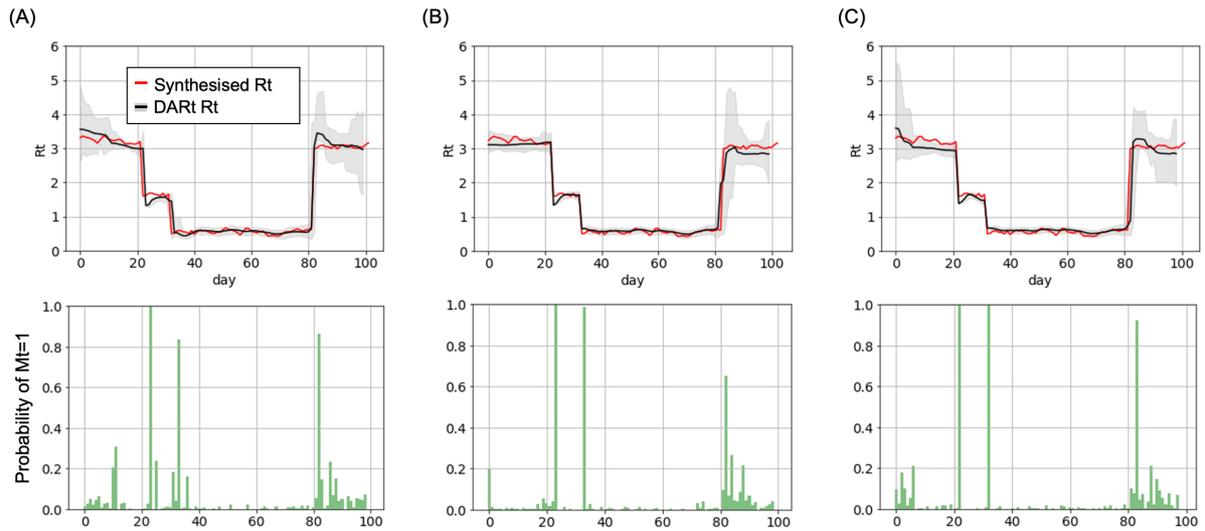

**The Supplementary Figure 7.** The $R_t$ estimation results of DARt with different truncation threshold: A) 0.01, B) 0.05 and C) 0.1.

### 6.3 Uncertainty in the delay distributions

In this paper, we focus on dealing with uncertainties brought by noisy observations. However, we are also quite interested in how our model would behave when uncertainties in the distributions of generation time and observation delay exist. Figure 8 A) shows the $R_t$ estimation results when the parameters of the generation time and observation delay distributions are set as follows: the distribution of generation time ~ Gamma(shape, scale), where $shape \sim \mathcal{N}(5.51, (10\% * 5.51)^2)$ and $scale \sim \mathcal{N}(0.81, (10\% * 0.81)^2)$; the distribution of observation delay distribution ~ Lognormal(mean, SD), where $mean \sim \mathcal{N}(1.64, (10\% * 1.64)^2)$ and $SD \sim \mathcal{N}(0.363, (10\% * 0.363)^2)$. We can see that the $R_t$ estimation results from uncertain time distributions [2] are still consistent with the synthesised $R_t$.

---

[2] The R function 'bootstrapped_dist_fit' used for fitting lognormal distribution is adopted from the EpiNow2 package.

Figure 8 B) further investigates the impact of choosing different forms of time distributions. Rather than assuming the generation time distribution follows a Gamma distribution, here we set it to be Lognormal. The parameters of the Lognormal distribution are obtained from fitting the original Gamma distribution by the Lognormal distribution[2]. The generation time distribution is hence set to be drawn from Lognormal(mean, SD), where $mean \sim \mathcal{N}(1.26, 0.042^2)$ and $SD \sim \mathcal{N}(0.46, 0.032^2)$. We can find that the estimated $R_t$ results at the beginning and ending time points are lower than the synthesised $R_t$. This is because the generation time distribution used for estimation is different from the original distribution used in the synthetic data generation. However, the general trend of $R_t$ with three abrupted changes is still well captured by DARt.

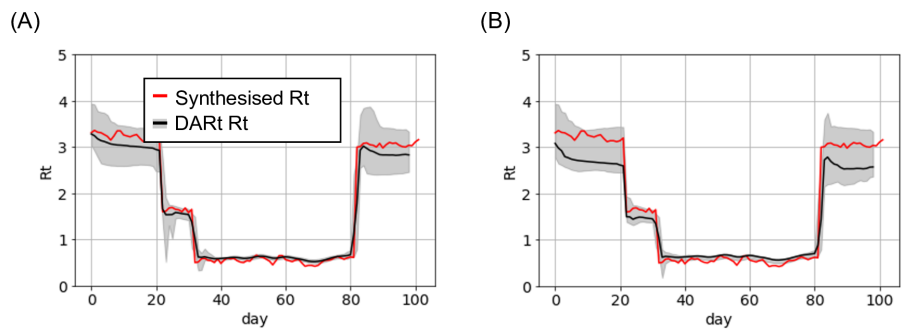

**The Supplementary Figure 8.** A) The $R_t$ estimation results of DARt obtained from the generation time and observation delay distributions with uncertainties. B) The $R_t$ estimation results of DARt obtained from the generation time distribution following a Lognormal distribution.

## 7 Supplementary table for simulation results

The simulation results in Figure 3 from the main manuscript have illustrated the performance of DARt. As indicated in the Supplementary Table 2, the simulation results further quantitively confirms our findings by calculating the mean and standard deviation of estimation differences over time. We can easily observe that: 1) compared with EpiNow2 and EpiEstim, DARt achieves much smaller estimation errors; 2) compared with DARt without smoothing, the full implementation of DARt with smoothing can reduce estimation errors, showing that smoothing has greatly contributed to uncertainty reduction.

**The Supplementary Table 2.** Simulation results using data synthesized in the main manuscript. $R_t$-mean/$J_t$-mean and Rt-sd/$J_t$-sd are the mean and standard deviation of the differences between synthesized $R_t$/$J_t$ and estimated $R_t$/$J_t$.

|  | $R_t$-mean | $R_t$-sd | $J_t$-mean | $J_t$-sd |
|---|---|---|---|---|
| EpiNow2 | 0.30 | 0.30 | 588.98 | 1283.08 |
| EpiEstim | 0.45 | 0.66 | NA | NA |
| DARt without smoothing | 0.21 | 0.33 | 729.42 | 2301.83 |
| DARt | 0.12 | 0.14 | 333.83 | 681.07 |